\documentclass{amsart}

\usepackage{amsfonts}
\usepackage[foot]{amsaddr}
\usepackage{graphicx}
\usepackage{caption}
\usepackage{subcaption}
\usepackage{url}

\newtheorem{rem}{Remark}

\graphicspath{{./images/}}

\begin{document}

\title[Adaptive Parallel-In-Space-Time Finite Element]{Massively parallel-in-space-time, adaptive finite element framework for non-linear parabolic equations}
%\headers{Adaptive Parallel-In-Space-Time Finite Element}{R. Dyja, B. Ganapathysubramanian, and K. van der Zee}

% Authors: full names plus addresses.
\author[R. Dyja]{Robert Dyja}
\address{Czestochowa University of Technology, ul. Dabrowskiego 73, 42-201 Czestochowa, Poland}
\email{robert.dyja@icis.pcz.pl}

\author[B. Ganapathysubramanian]{Baskar Ganapathysubramanian}
\address{Iowa State University, 306 Lab of Mechanics, Ames, IA 50011}
\thanks{We gratefully acknowledge NSF~1435587 and XSEDE resources at TACC, as well as an exploratory account on BlueWaters.}
\email{baskarg@iastate.edu}

\author[K.G. van~der~Zee]{Kristoffer G. van~der~Zee}
\address{University of Nottingham, School of Mathematical Sciences, University Park, Nottingham NG7~2RD, UK}
\thanks{Kristoffer G. van~der~Zee's research was supported by the Engineering and Physical Sciences Research Council (EPSRC) under grant EP/I036427/1}
\email{KG.vanderZee@nottingham.ac.uk}


\begin{abstract}
%method
We present an adaptive methodology for the solution of (linear and) non-linear time dependent problems that is especially tailored for massively parallel computations. The basic concept is to solve for large blocks of space-time unknowns instead of marching sequentially in time. The methodology is a combination of a computationally efficient implementation of a parallel-in-space-time finite element solver coupled with {\it a posteriori} space-time error estimates and a parallel mesh generator. This methodology enables, in principle, simultaneous adaptivity in both space and time (within the block) domains. We explore this basic concept in the context of a variety of time-steppers including $\Theta$-schemes and Backward Differentiate Formulas. We specifically illustrate this framework with applications involving time dependent linear, quasi-linear and semi-linear diffusion equations. We focus on investigating how the coupled space-time refinement indicators for this class of problems affect spatial adaptivity. Finally, we show good scaling behavior up to 150,000 processors on the Blue Waters machine. This is achieved by careful usage of memory via block storage and non-zero formats along with lumped communication for matrix assembly. This methodology enables scaling on next generation multi-core machines by simultaneously solving for large number of time-steps, and reduces computational overhead by refining spatial blocks that can track localized features. This methodology also opens up the possibility of efficiently incorporating adjoint equations for error estimators and inverse design problems, since blocks of space-time are simultaneously solved and stored in memory.  
\end{abstract}

\maketitle

% REQUIRED
%\begin{keywords}
%  parabolic problems, parallel-in-time, finite element method, adaptive mesh refinement
%\end{keywords}

% REQUIRED
%\begin{AMS}
%  65M22, 65Y05
%\end{AMS}

\section{Introduction}
We describe the methodology, implementation details and application examples of space-time block adaptive solutions to parabolic partial differential equations (evolution problems). This approach is primarily motivated by the necessity of designing computational methodologies that can scale to leverage the availability of very large computing clusters (with $\sim$ 10's to 100's of thousands of processors). For evolution problems, the standard approach of decomposing the spatial domain is a powerful paradigm of parallelization. However, for a fixed spatial discretization, the efficiency of purely spatial domain decomposition degrades substantially beyond a threshold (usually $\sim$ thousands of processors, beyond which efficiency is communication limited) which make this approach unsuitable on larger machines. To overcome this barrier, a natural approach is to consider the time domain as an additional dimension and simultaneously solve for blocks of time, instead of the standard approach of sequential time-stepping. Early work on this approach was considered by Hughes and coworkers~\cite{hughes_hulbert, hughes_steward}, Tezduyar and co-workers~\cite{tezduyar},  and Reddy and co-workers~\cite{potanza_reddy}, while variations on this theme have recently been explored by several groups~\cite{behr, carey_estep, lowrie1996, mani2011, rendall2012, wang2015}. Other examples include methods that exploit the fact that some coefficients during matrix assembly can be calculated concurrently~\cite{burrage} and methods that transform the equations so as to build one system of equations for more than one timestep~\cite{parareal2001}.

\par The concept of solving for blocks of time simultaneously has recently gained a lot of attention to enable effective usage of exascale computing resources (see for instance the US DOE's Exascale Mathematics Working Group \cite{siam2013}). In addition to this obvious advantage, solving for space-time blocks also allows natural incorporation of {\it a posteriori} error estimates for mesh adaptivity, and enables the solution of inverse problems (involving adjoints). This has several additional tangible benefits in the context of computational overhead. For evolution problems that exhibit “regionalized” behavior in space and time (wave equations, moving interfaces like bubbles and shocks) solving in blocks of space-time that are locally refined to match the regional behavior provides substantial computational gain~\cite{carey_estep}. Similarly, the availability of error estimates across a block of time allows optimal choices of space and time adaptivity. 

\par Motivated by these considerations, this paper presents a methodology for the solution of linear, quasi-linear and semi-linear (time dependent) diffusion equations in three dimensions. We discuss the development of a parallel adaptive framework for the solution of large blocks of space-time. We detail the development of the block space-time framework for two classes of time-steppers ($\theta$ schemes, Backward Difference Formula (BDF)). We discuss implementation details of the massively parallel adaptive block-space time framework and illustrate scaling behavior up to 150,000 processors. We subsequently define {\it a posteriori space-time error indicators} to identify spatial regions for mesh adaption. Finally, we demonstrate that for sufficiently large problems the block space-time approach is much more computationally efficient than the standard sequential time-stepping approach. 

\par The outline of the rest of the paper is as follows: Section~2, and~3 detail the block space-time framework for linear, and non-linear evolution equations, respectively. Section~4 discusses the space-time error estimates for these classes of problems. In Section~5, we discuss implementation details and show scaling performance and analysis. Section~6 illustrates several numerical examples of the framework. We conclude in Section~7.

\section{Basic space-time formulation: linear and non-linear versions}

\subsection{Space-time framework for a linear problem}

\noindent Given a bounded domain $\Omega \in R^3$, and a finite time domain $\left[ 0, T\right]$, consider the parabolic equation that solves for $u \colon \Omega \times \left[ 0, T\right]\rightarrow \mathbb{R}$:
\begin{equation}
	\begin{cases}
	\partial_t u(\mathbf{x},t) - \nabla \cdot \kappa \nabla u(\mathbf{x},t) = f(\mathbf{x}, t) \quad \text{in } \Omega \times \left[ 0, T\right]\\
	u(\mathbf{x},0) = u_0
	\end{cases}
	\label{poisson}
\end{equation}
where $f \colon \Omega \times \left[ 0, T\right] \rightarrow \mathbb{R}$ is a smooth source function, and $\kappa > 0$. We consider (without loss of generality) that Dirichlet boundary conditions are imposed on the boundary $\Gamma$ , unless otherwise specified. Considering a tesselation, $\mathcal{T} \equiv \{ \Omega^1,...,\Omega^e,...\}$, of the domain $\Omega$ into elements (with average size $h$), the weak form of this equation is given as:
\begin{equation}
    \begin{cases}
	\text{Find } & u^h(\cdot, t) \in \mathcal{U}^h: \\
	 & (w^h, \partial_t u^h(\cdot, t)) + (\nabla w^h, \kappa \nabla u^h(\cdot, t)) = (w^h, f) \quad \forall w^h \in \mathcal{V}^h
    \end{cases}
\end{equation}
where $\left( \ldotp , \ldotp \right)$ is the $L_2$ inner product on $\Omega$ and
\begin{equation}
    \begin{gathered}
	\mathcal{U}^h := \left\{ u^h | u^h \in H^1(\Omega), \quad u^h \in P(\Omega^e) \quad \forall e \right\} \\
	\mathcal{V}^h := \left\{ w^h | w^h \in H^1(\Omega), \quad w^h \in P(\Omega^e) \quad \forall e \right\}
    \end{gathered}
\end{equation}
with $P(\Omega^e)$ being the space of the standard polynomial finite element shape functions on element $\Omega^e$. To obtain a fully discretized form, we employ a time stepping technique on the above semi-discrete equation. While any time-stepping method can be used, as an example, consider the Euler Backward Formula that is defined on a discretization $ \{0, t_1, t_2,...,T\} $ of the time domain:
\begin{equation}
\left( w^h, \frac{u_{n+1}^h - u_{n}^h}{\Delta t} \right) - 
\left(  \nabla w^h, \kappa \nabla u_{n+1}^h \right) = \left( w^h, f_{n+1} \right),~~ \text{for}~n=0,1,...
\label{EB_iter}
\end{equation}
where the subscript denotes evaluation at that discrete time, and $\Delta t = t_{n+1}-t_{n}$ is the timestep. 

Following standard FEM practice, with the tesselation of the domain resulting in $k$ nodal values (to describe spatial variation) of  the field $u$, equation~\eqref{EB_iter} can be expressed in terms of matrix-vector products as:
\begin{equation}
\mathbf{M}\mathbf{u}_{n+1} - \Delta t \mathbf{K}\mathbf{u}_{n+1} = 
\mathbf{M}\mathbf{u}_{n} + \Delta t \mathbf{f}_{n+1},~~ \text{for}~n=0,1,...
\label{EB_iter_matrix}
\end{equation}
where $\mathbf{M}$ and $\mathbf{K}$ are the global mass and stiffness matrices, respectively.\footnote{With some abuse of notation, the elements of matrix $\mathbf{M}$ are equal to $M_{ij} = (w_i^h, w_j^h)$ and matrix $\mathbf{K}$ are equal to $K_{ij} = (\nabla w_i^h,\kappa \nabla w_j^h)$.} $\mathbf{u}_{n+1}$ and $\mathbf{u}_{n}$ are vectors containing the nodal values of the field $u$ at time step $n+1$ and $n$, respectively. Equation~\eqref{EB_iter_matrix} represents the system of equations solved to get the solution for timestep $n+1$. The size of vector $\mathbf{u}$ is equal to the number of nodal unknowns, $k$. Similarly matrices $\mathbf{K}$, $\mathbf{M}$ are sparse matrices of size $k \times k$.

\begin{figure}[bth!]
\begin{center}
    \includegraphics[width=0.6\textwidth]{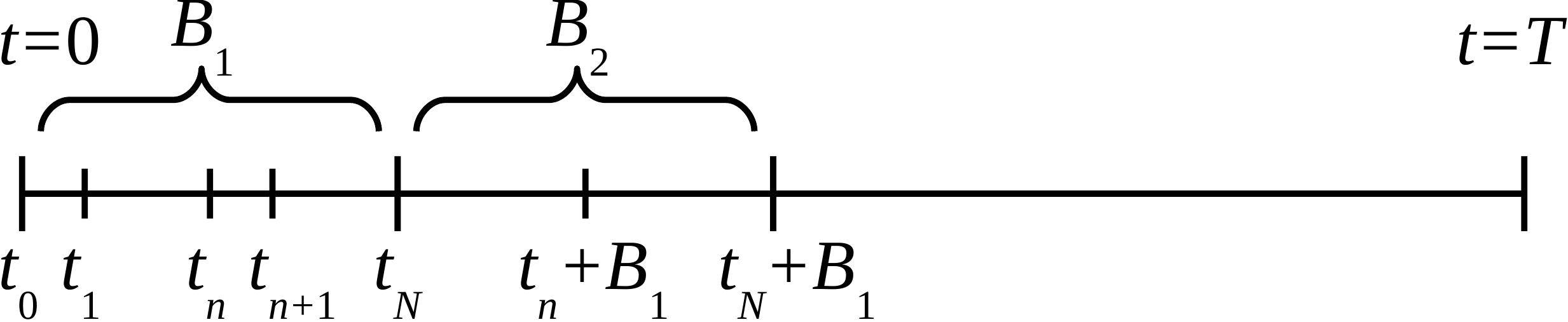}
    \caption{\label{timeblock} Indexing of timesteps per block (B). Number of timesteps per block is equal to N.}
\end{center}
\end{figure}

Consider a block-wise division of the total time domain. Each block, $B_i$, consists of multiple timesteps. This is schematically represented in Figure~\ref{timeblock}. Instead of sequentially solving for each time step (as in~\eqref{EB_iter_matrix}), consider solving for the field variable in a complete time block, $B$, consisting of $N$ timesteps simultaneously, i.e. solve for $\mathbf{u}_i,~ i=1,...,N$ simultaneously. 

This results in a block diagonal matrix  of size $(N \times k)$-by-$(N \times k)$ given by:
\begin{equation}
    \begin{bmatrix}
    \mathbf{I} & & & \\
	 -\mathbf{M} & \mathbf{M} + \Delta t \mathbf{K} & & \\
	 & -\mathbf{M} & \mathbf{M} + \Delta t \mathbf{K} & \\
	 & & \ddots & \\
	 & & -\mathbf{M} & \mathbf{M} + \Delta t \mathbf{K}
    \end{bmatrix}
    \begin{bmatrix}
    \mathbf{u}_{0} \\
	\mathbf{u}_{1} \\
	\mathbf{u}_{2} \\
	\vdots \\
	\mathbf{u}_{N}
    \end{bmatrix}
=
    \begin{bmatrix}
    \mathrm{IC} \\
	\Delta t \mathbf{f}_{1} \\
	\Delta t \mathbf{f}_{2} \\
	\vdots \\
	\Delta t \mathbf{f}_{N}
    \end{bmatrix}
    \label{EB_spacetime_matrix}
\end{equation}
where $\mathbf{I}$ is an identity matrix (of size $k$) and $\mathrm{IC}$ are the imposed initial conditions. This system solves for $N$ timesteps at once with a total number of unknowns $N \times k$. 

\begin{rem}
By treating the (unknown) nodal values at different time steps as multiple degrees of freedom associated with each spatial node, we can leverage standard algorithmic approaches (assembly, memory usage) tailored for multiple d.o.f problems. Many approaches in uncertainty quantification (polynomial chaos representation, spectral stochastic methods) leverage such an approach of representing field variation along additional dimensions (stochastic dimensions) as simply additional degrees of freedom at each spatial location \cite{narayan, wan, debusschere, soize}.
Figure~\ref{linear_entries}(left) illustrates an elemental matrix considering $N (=10)$ time steps (i.e. $10$ degrees of freedom per node), for a 2D quadrilateral element \footnote{using Backward-Euler discretization in time}. Note the sparse structure of the elemental matrix as well as the resulting global matrix (for a $2 \times 2$ quad mesh) in Figure~\ref{linear_entries} (right). 

\begin{figure}[bth!]
\begin{center}
	\includegraphics[width=0.35\textwidth]{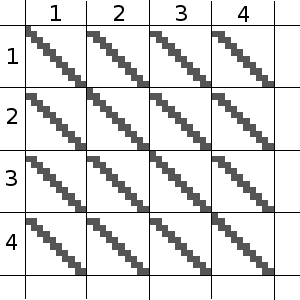}
	\includegraphics[width=0.35\textwidth]{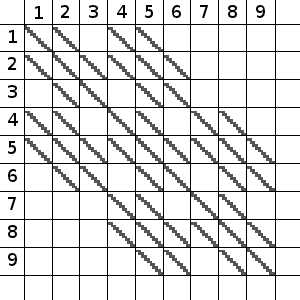}
    \caption{\label{linear_entries} Left, Nonzero entries in an elemental matrix. Right,  Nonzero entries in the resultant global matrix for a 2x2 mesh of quadrilateral elements. Dark squares represent nonzero entries and number are node indices. The time block consists of 10 time steps. }
\end{center}
\end{figure}
\end{rem}

\subsection{Space-time framework for non-linear problems}
Extending the approach to certain non-linear problems is straightforward. Consider the case where $\kappa$ is a function of  the dependent variable, $u$. We assume that $\kappa(u)$ satisfies appropriate smoothness and boundedness assumptions to ensure existence and uniqueness, $0 < \underline{\kappa} \leq \kappa(u) \leq \overline{\kappa} < \infty $.  In this case the weak form for a  block $B$ is 
\begin{equation}
	(w^h, \frac{u^h_{n+1} - u^h_{n}}{\Delta t}) + (\nabla w^h, \kappa(u^h_{n+1}) \nabla u^h_{n+1}) = (w^h, f_{n+1}),~~ \text{for}~n=0,1,...,N
\label{eq_weak_nl}
\end{equation}
The solution to such non-linear equations are usually via (quasi-)Newton schemes. The methodology involves construction of the Jacobian  and residual, which are used to compute updates. This is represented in matrix-vector terms as: 
\begin{eqnarray}
	\mathbf{J}_{\mathbf{u}_{n+1}^{i}} \delta \mathbf{u}_{n+1}^{i+1} = \mathbf{F}_{\mathbf{u}_{n+1}^{i}},
	\quad
	\mathbf{u}_{n+1}^{i+1} = \mathbf{u}_{n+1}^{i} + \delta \mathbf{u}_{n+1}^{i+1}, \\ \nonumber ~~ \text{for}~i=1,...,\text{until convergence} ~~ \text{and for}~n=0,1,...
\label{eq:nl_seq}
\end{eqnarray}
where $\mathbf{J}_{\mathbf{u}_{n+1}^{i}}$ is the Jacobian (or linearized form), and $\mathbf{F}_{\mathbf{u}_{n+1}^{i}}$ is the residual of the above equation, both computed using $\mathbf{u}_{n+1}^{i}$. More specifically,  for the non-linear diffusion equations defined by~\eqref{eq_weak_nl}, the residual, $\mathbf{F}_{\mathbf{u}_{n+1}^{i}}$, is given by 
\begin{equation}
	\mathbf{F}_{\mathbf{u}_{n+1}^{i}} = \frac{1}{\Delta t}\mathbf{M}\mathbf{u}_{n+1}^{i} - \frac{1}{\Delta t}\mathbf{M}\mathbf{u}_{n} - \mathbf{K}(\mathbf{u}_{n+1}^{i})\mathbf{u}_{n+1}^{i} -  \mathbf{f}_{n+1}
\end{equation}
where $\mathbf{K}(\mathbf{u}_{n+1}^{i})$ denotes the solution dependent stiffness matrix. The Jacobian is given as:
\begin{equation}
	\mathbf{J}_{\mathbf{u}_{n+1}^{i}} = \frac{1}{\Delta t}\mathbf{M} + \left( \mathbf{K}(\mathbf{u}_{n+1}^{i}) + \frac{\mathrm{d}\mathbf{K}(\mathbf{u}_{n+1}^{i})}{\mathrm{d}u} \right)
\end{equation}
 
Instead of sequentially solving for each time step (as in~\eqref{eq:nl_seq}), consider solving for the field variable in a complete time block, $B$, consisting of $N$ timesteps simultaneously. That is,

\begin{multline}
    \begin{bmatrix}
	\mathbf{J}_{\mathbf{u}_{1}^{i}} & & & \\
	 - \frac{1}{\Delta t}\mathbf{M}& \mathbf{J}_{\mathbf{u}_{2}^{i}} & & \\
	 &  & \ddots & \\
	  &  &  -\frac{1}{\Delta t}\mathbf{M} & \mathbf{J}_{\mathbf{u}_{N}^{i}}
    \end{bmatrix}
    \begin{bmatrix}
	\delta \mathbf{u}^{i+1}_{1} \\
	\delta \mathbf{u}^{i+1}_{2} \\
	\vdots \\
	\delta \mathbf{u}^{i+1}_{N}
    \end{bmatrix}
=
    \begin{bmatrix}
	\mathbf{F}_{\mathbf{u}_{1}^{i}} \\
	\mathbf{F}_{\mathbf{u}_{2}^{i}} \\
	\vdots \\
	\mathbf{F}_{\mathbf{u}_{N}^{i}}
    \end{bmatrix}
    \\
~~ \text{and}~~
    \\
    \begin{bmatrix}
	\mathbf{u}^{i+1}_{1} \\
	\mathbf{u}^{i+1}_{2} \\
	\vdots \\
	\mathbf{u}^{i+1}_{N}
    \end{bmatrix}
=
    \begin{bmatrix}
	\mathbf{u}^{i}_{1} \\
	\mathbf{u}^{i}_{2} \\
	\vdots \\
	\mathbf{u}^{i}_{N}
    \end{bmatrix}
+
    \begin{bmatrix}
	\delta \mathbf{u}^{i+1}_{1} \\
	\delta \mathbf{u}^{i+1}_{2} \\
	\vdots \\
	\delta \mathbf{u}^{i+1}_{N}
    \end{bmatrix}
\end{multline}

\begin{rem}
One can alternatively ignore the off-diagonal entries of the block Jacobian to construct an approximate diagonal Jacobian. The propagation of time information is then limited to the residual on the right hand side. We tested both approaches, with the latter approach taking marginally more iterations to convergence, while providing substantial ease of implementation. Unless otherwise stated, all our results are based on the latter approach.
\end{rem}

\section{Space-time formulation: Higher Order Time schemes}

We next look at extending the space-time strategy to incorporate two families of higher order time-steppers: $\Theta$ scheme and Backward difference formula. We consider linear and nonlinear diffusion, and moreover, we also consider the treatment of the Allen-Cahn equation, which is a parabolic PDE with a lower-order nonlinearity, whose solution has evolving layers and for which adaptivity is particularly useful.
\subsection{$\Theta$ scheme: Linear equation}

\noindent The semi-discrete form of the $\Theta$-scheme -- which is a generalization of the Euler Backward scheme -- is as follows:
\begin{multline}
	(w, u_{n + 1}) - (w, u_{n}) + \Delta t \left[ (1 - \Theta) (\nabla w, \kappa \nabla u_{n}) + \Theta (\nabla w, \kappa \nabla u_{n + 1}) \right] = {} \\
	{} \Delta t \left[ (1 - \Theta) (w, f_{n}) + \Theta (w, f_{n+1}) \right],~~ \text{for}~n=0,1,...
\end{multline}

\noindent The fully discrete matrix-vector representation is given by
\begin{equation}
\mathbf{M}\mathbf{u}_{n+1} - \mathbf{M}\mathbf{u}_{n} + \Delta t (1-\Theta)\mathbf{K}\mathbf{u}_{n} + \Delta t \Theta \mathbf{K}\mathbf{u}_{n+1} = 
\Delta t (1-\Theta) \mathbf{f}_{n}+\Delta t \Theta \mathbf{f}_{n+1},~~ \text{for}~n=0,1,...
\label{theta_iter_matrix}
\end{equation}

\noindent Again, it is straightforward to group and simultaneously solve for $N$ time steps together.  The corresponding matrix form is expressed as:
\begin{eqnarray}
\nonumber
    \begin{bmatrix}
    \mathbf{I} & & & \\
	-\mathbf{M} + \Delta t (1 - \Theta) \mathbf{K} & \mathbf{M} + \Delta t \Theta \mathbf{K} & & \\
	 & -\mathbf{M} + \Delta t (1 - \Theta) \mathbf{K} & \mathbf{M} + \Delta t \Theta \mathbf{K} & \\
	 & & \ddots & \\
	 & & -\mathbf{M} + \Delta t (1 - \Theta) \mathbf{K} & \mathbf{M} + \Delta t \Theta \mathbf{K}
    \end{bmatrix}
    \begin{bmatrix}
    \mathbf{u}_{0} \\
	\mathbf{u}_{1} \\
	\mathbf{u}_{2} \\
	\vdots \\
	\mathbf{u}_{N}
    \end{bmatrix} \\ 
=
    \begin{bmatrix}
    \mathrm{IC} \\
	\Delta t \left[ (1 - \Theta) \mathbf{f}_{0} + \Theta \mathbf{f}_{1} \right] \\
	\Delta t \left[ (1 - \Theta) \mathbf{f}_{1} + \Theta \mathbf{f}_{2} \right] \\
	\vdots \\
	\Delta t \left[ (1 - \Theta) \mathbf{f}_{N-1} + \Theta \mathbf{f}_{N} \right]
    \end{bmatrix}
\label{linear_system}
\end{eqnarray}
Again, the global space-time matrix (\ref{linear_system}) has a block structure.

\subsection{$\Theta$ scheme: Nonlinear diffusion with variable coefficient}
The corresponding weak form for this case is given as:
\begin{multline}
	(w^h, \frac{u^h_{n+1} - u^h_{n}}{\Delta t}) +(1- \Theta) (\nabla w^h, \kappa(u^h_{n}) \nabla u^h_{n}) +\Theta (\nabla w^h, \kappa(u^h_{n+1}) \nabla u^h_{n+1}) = \\
	(1-\Theta) (w^h, f_{n})+ \Theta(w^h, f_{n+1}),~~ \text{for}~n=0,1,...
\label{eq_weak_nl_var}
\end{multline}
The Jacobian is: 
\begin{equation}
	\mathbf{J}_{\mathbf{u}_{n+1}^{i}}
=	\mathbf{M} + \Delta t \Theta \left( \mathbf{K}(\mathbf{u}_{n+1}^{i}) + \frac{\mathrm{d}\mathbf{K}(\mathbf{u}_{n+1}^{i})}{\mathrm{d}u} \right)
\end{equation}
while the residual is given by:
\begin{equation}
	\mathbf{F}_{\mathbf{u}_{n+1}^{i}}
=
	\mathbf{M}\mathbf{u}^i_{n+1} - \mathbf{M}\mathbf{u}_{n}
	+ \Delta t \Theta \left( \mathbf{K}(\mathbf{u}^i_{n+1})\mathbf{u}^i_{n+1} - \mathbf{f}_{n+1} \right) 
	+ \Delta t (1- \Theta) \left( \mathbf{K}(\mathbf{u}_{n})\mathbf{u}_{n} - \mathbf{f}_{n} \right)
\end{equation}

\noindent Again, it is straightforward to group and simultaneously solve for $N$ time steps together.  The corresponding (diagonal) block Jacobian (see Remark 2) is given as:
\begin{equation}
    \begin{bmatrix}
	\mathbf{M} + {} &  &\\
	 \Delta t \Theta \left( \mathbf{K}(\mathbf{u}_{1}^{i}) + \frac{\mathrm{d}\mathbf{K}(\mathbf{u}_{1}^{i})}{\mathrm{d}u} \right) &  &\\
	 & \mathbf{M} + {} &  &\\
	 &  \Delta t \Theta \left( \mathbf{K}(\mathbf{u}_{2}^{i}) + \frac{\mathrm{d}\mathbf{K}(\mathbf{u}_{2}^{i})}{\mathrm{d}u} \right) &  &\\
	 & \ddots  & & \\
	 & & \mathbf{M} + {}  & \\
	 & & \Delta t \Theta \left( \mathbf{K}(\mathbf{u}_{N}^{i}) + \frac{\mathrm{d}\mathbf{K}(\mathbf{u}_{N}^{i})}{\mathrm{d}u} \right) &
    \end{bmatrix}
\end{equation}
As before, the upper index denotes Newton-Raphson iteration. 
\subsection{$\Theta$ scheme: Allen Cahn Equation}
The Allen-Cahn equation is a semi-linear diffusion equation with a non-linear reaction term:

\begin{equation}
	\partial_t u(\mathbf{x},t) - \Delta u(\mathbf{x},t) + \epsilon^{-2} f(u) = 0 
\end{equation}
where we set $f(u)=u(u^2 - 1)$. The initial condition is $u(\mathbf{x},0) = u_0 $ along with zero flux conditions in the boundaries.

\noindent The corresponding semi-discrete form is given as:
\begin{equation}
	(w^h, \partial_t u^h) + (\nabla w^h, \nabla u^h) + \epsilon^{-2} (w^h, f(u^h)) = 0
\end{equation}

\noindent Using the $\Theta$-scheme results in the fully discrete form:
\begin{multline}
	(w^h, u^h_{n+1}) + \Delta t \Theta \left [ (\nabla w^h, \nabla u^h_{n+1})+ \epsilon^{-2} (w^h, f(u^h_{n+1}))\right] = \\
	(w^h, u^h_{n})  - \Delta t (1 - \Theta) \left[ (\nabla w^h,\nabla u^h_{n}) + \epsilon^{-2} (w^h, f(u^h_{n})) \right] 
\end{multline}

\noindent Again, it is straightforward to group and simultaneously solve for $N$ time steps together. The corresponding (diagonal) block Jacobian is given as:
\begin{equation}
    \begin{bmatrix}
	\mathbf{M} + \Delta t \Theta \left( \mathbf{K}(\mathbf{u}_{1}^{i}) +  \right. & & & \\
	 \left. \epsilon^{-2}  \frac{\mathrm{d}\mathbf{f}(\mathbf{u}_{1}^{i})}{\mathrm{d}u} \right) & & & \\
	 & \mathbf{M} + \Delta t \Theta \left( \mathbf{K}(\mathbf{u}_{2}^{i}) +  \right.  & \\
	 & \left. \epsilon^{-2}  \frac{\mathrm{d}\mathbf{f}(\mathbf{u}_{2}^{i})}{\mathrm{d}u} \right)  & \\
	 & & \ddots & \\
	 & & & \mathbf{M} + \Delta t \Theta \left( \mathbf{K}(\mathbf{u}_{N}^{i}) + \right. \\
	 & & & \left. \epsilon^{-2}  \frac{\mathrm{d}\mathbf{f}(\mathbf{u}_{N}^{i})}{\mathrm{d}u} \right)
    \end{bmatrix}
\end{equation}
Alternative schemes for the Allen-Cahn equation and other phase-field models are described in, e.g., \cite{gomez_zee}.

\subsection{Backward difference formula based time steppers} 
\begin{figure}[bth!]
\begin{center}
    \includegraphics[width=0.35\textwidth]{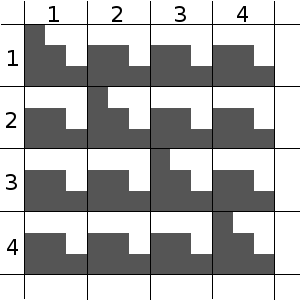}
	\includegraphics[width=0.35\textwidth]{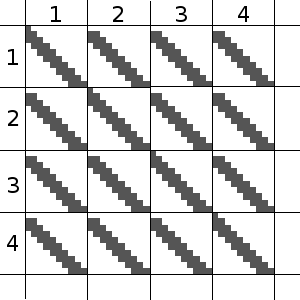}
    \caption{\label{bdf_entries} View of nonzero entries of an elemental matrix (2D quad with linear basis functions) for a block of space-time consisting of 2 timesteps + initial condition (left) and 9 timesteps + initial condition(right) with Backward difference formula of second order. First timestep is calculated using Euler Backward scheme.}
\end{center}
\end{figure}

BDF based time-steppers of order $s$ utilize the solution at $s$ previous time steps to construct the solution at the next time step. A general $s$ order BDF scheme is given as:
\begin{equation}
	\sum \limits_{k=0}^{s} \alpha_{k} u_{n+k} = \Delta t \beta g_{n+s}
\end{equation}
where the left hand side is the BDF scheme representation of the time derivative, $\frac{\partial u}{\partial t}$, in terms of the solution $u_i$ at time point $i$, and the right hand side collects all other terms.  Here, $\alpha$ and $\beta$ are known BDF coefficients \cite{butcher}. A first order ($s=1$) BDF scheme is identical to the Euler Backward scheme described earlier. The simplest multi-step scheme is for $s=2$ and for the linear diffusion equation it is given as:
\begin{equation}
	(w^h, u^h_{n+2}) - \frac{4}{3}(w^h, u^h_{n+1}) + \frac{1}{3}(w^h, u^h_n) + \frac{2}{3}\Delta t (\nabla w^h, \kappa\nabla u^h_{n+2})  = \frac{2}{3}\Delta t (w^h, f_{n+2})
	\label{bdf2_example}
\end{equation}
Again, it is straightforward to group and simultaneously solve for $N$ time steps together.\footnote{Note that the first time step is approximated using a backward Euler time stepper as the second order BDF scheme requires knowledge of the solution at two previous time steps} The corresponding space-time block equations are as follows:
\begin{equation}
    \begin{bmatrix}
    \mathbf{I} & & & & \\
	-\mathbf{M} & \mathbf{M} + \Delta t \mathbf{K} & & & \\
	\frac{1}{3}\mathbf{M} & -\frac{4}{3}\mathbf{M} & \mathbf{M} + \frac{2}{3} \Delta t \mathbf{K} & & \\
	 & & & \ddots & \\
	 & & \frac{1}{3} \mathbf{M} & -\frac{4}{3} \mathbf{M} & \mathbf{M} + \frac{2}{3} \Delta t \mathbf{K}
    \end{bmatrix}
    \begin{bmatrix}
    \mathbf{u}_{0} \\
	\mathbf{u}_{1} \\
	\mathbf{u}_{2} \\
	\vdots \\
	\mathbf{u}_{N}
    \end{bmatrix}
=
    \begin{bmatrix}
    \mathrm{IC} \\
	\Delta t \mathbf{f}_{1} \\
	\frac{2}{3} \Delta t \mathbf{f}_{2} \\
	\vdots \\
	\frac{2}{3} \Delta t \mathbf{f}_{N}
    \end{bmatrix}
\label{bdf_linear_system}
\end{equation}
It is clear that higher order multistep methods produce block matrices that have a larger bandwidth (see Figure \ref{bdf_entries}). Nonlinear problems can be treated similarly.

\section{Adaptive meshing for the block-space-time method: Residual based error estimator}
A central idea of this work is to develop a block space-time methodology that can be integrated with mesh adaptivity. This will enable targeted refinement of regions that exhibit variations in the corresponding block of time. Mesh adaptivity requires the definition of an indicator function that determines which regions of the space require refinement/coarsening. In this work, we build on prior work and utilize standard residual-based error indicators; see, e.g. \cite{ainsworth_oden, verfurth2013, verfurth1996}. Alternative duality-based indicators for nonlinear parabolic problems are described in, e.g. \cite{carey_estep, simsek_zee, eriksson_johnson}.

Residual based error indicators $\eta_e$ are constructed for each spatial element, $\Omega_e$, and consist of two terms -- an interior residual, $r_{int}$, and a jump residual, $r_{jump}$~\cite{verfurth2013}:
\begin{equation}
	\eta_e^2 = h_e^2 \| r_{int}^h \|_{L^2(\Omega_e)}^2 + h_e \| r_{jump}^h \|_{L^2(\partial \Omega_e)}^2 \label{error_indicator}
\end{equation}
where $h_e$ is the size of element, $\Omega_e$. For example, for the linear and quasi-linear diffusion equation, the detailed derivation of the interior and jump residual is available in the work of Verfurth~\cite{verfurth2013}. We refer the interested reader to that work and only show the key results here. Basically, the two terms are constructed (as the name suggests) from the definition of the residual:
\begin{equation}
\mathcal{R}^h(w) = (w, f) - (w, \partial_t u^h) - (\nabla w, \kappa(u^h) \nabla u^h)
\end{equation}
This residual is decomposed into element-wise terms as:
\begin{equation}
R^h(w) =
\sum \limits_e \left(
\int_{\Omega_e} w r_{\text{int}}^h \, \mathrm{d} \Omega  +
\int_{\Gamma_e} w r_{\text{jump}}^h \, \mathrm{d} \Gamma
\right)
\end{equation}
where
\begin{equation}
r_{\text{int}}^h = f - \partial_t u^h + \nabla \cdot \kappa(u^h) \nabla u^h
\end{equation}
and
\begin{equation}
r_{\text{jump}}^h = 
\left\{ \begin{array}{lcl}
0 & \text{on} & \partial \Omega_e \cap \Gamma_h  \\
\kappa(u^h) \nabla u^h_+ \cdot \mathbf{n}_+ + \kappa(u^h) \nabla u^h_- \cdot \mathbf{n}_- & \text{on} & \partial \Omega_e \setminus \Gamma_h
\end{array} \right.
\end{equation}
where $\Gamma_h$ is that part of the boundary with Dirichlet conditions imposed.

\noindent The interior and jump residuals for the Allen-Cahn equations are similarly defined as 
\begin{equation}
r_{\text{int}}^h = \epsilon^{-2} f(u^h) - \partial_t u^h + \Delta u^h
\end{equation}
and
\begin{equation}
r_{\text{jump}}^h = 
\left\{ \begin{array}{lcl}
0 & \text{on} & \partial \Omega_e \cap \Gamma_h  \\
\nabla u^h_+ \cdot \mathbf{n}_+ + \nabla u^h_- \cdot \mathbf{n}_- & \text{on} & \partial \Omega_e \setminus \Gamma_h
\end{array} \right.
\end{equation}

Recall that we are using error indicators defined over a block of time. We extend the concept of error indicator defined at one time step to the notion of an error indicator defined over a block of time steps. There are several choice of error indicators with two of them being: \\
the average value of the error indicator across the block of time
\begin{equation}
\eta_{e,avge} = \sum \limits_{\text{Block of time, B}}  \left( \frac{\sum_{n=1}^{N} \eta_{e, n}}{N} \right)
\label{eqn:error_indicator1}
\end{equation}
and the maximum value in the block of time:
\begin{equation}
\eta_{e,max} =     \max_{\text{Block of time, B}} \eta_{e, n} 
\end{equation}
The former approach is advantageous for slow variations in the block and avoids frequent migrating of elements between processors, which can negatively affect scalability. The latter approach is advantageous when there is rapid changes across a few timesteps. In this case the error estimator is  not 'diffused' by timesteps where solution is changing slowly.
\begin{rem}
In this work, we approximate the semi-discrete term, $\frac{\partial u}{\partial t}$ in terms of its finite difference representation ($\Theta$ or BDF scheme). An implicit assumption is that the time steps are small enough that this approximation is valid. Ideally, one would choose a consistent representation in both space and time; i.e. using a finite element representation for time variations~\cite{potanza_reddy,hughes_hulbert}. This provides several advantages in terms of mathematical elegance. We defer this development to a subsequent paper.
\end{rem}

\section{Implementation details}
We utilize our in-house scalable, parallel FEM framework that is optimized for distributed memory computing. The FEM software library is implemented in C++ and uses object oriented software principles. Linear algebra, parallel matrix and vector storage are all performed by the PETSc library~\cite{petsc}. PETSc modules (KSP, SNES) are used to solve (non)linear equations. The  FEM library is dynamically linked to the Parallel Hierarchical Grid (PHG) library~\cite{phg} which is a parallel mesh refinement framework. PHG uses a bisection type algorithm~\cite{zhang}, specifically newest vertex bisection to refine/coarsen elements\footnote{In newest vertex bisection, the edge that lies in opposite to the newest node is divided}. PHG operates on simplex elements and produces conforming meshes after refinement.
While our existing implementation was reasonably optimized, several software engineering principles had to be carefully implemented to ensure efficient execution of the block space-time  problems. Some of the simpler changes that had substantial impact for the space-time formulation\footnote{even very simple changes possibly like avoiding division operations by replacing with multiplication, or using 'const' to enable the compiler to automatically simplify expressions, and avoiding unnecessary conversions between integer and double values have some impact}, {\it but had minimal impact on the existing, standard iterative formulation}\footnote{these standard best practices provide minimal improvements to sequential time stepping, and are usually not reported or ignored.} are: (a) moving calculations out of loops whenever possible, (b) executing calculations and storing results in array before loops, (c) conversion to $64$-bit based integer variables for storing matrix indices, which allows going beyond 4 billion unknowns, (d) using local values instead of values obtained through pointer or reference\footnote{the function can access values more quickly, as it does not have to first fetch pointer, than check where pointer points and get value}.

We made software engineering decisions to ensure that the space-time implementation is compatible with our existing sequential FEM frameworks (see remark 1). Basically, the space-time formulation is treated as the solution of a steady-state problem. Initial conditions are imposed as boundary condition on the first degree of freedom.  We discuss key aspects of the implementation next.

\subsection{Memory interlacing and matrix bandwidth}
We rearrange the vector of unknowns $\mathbf{u}$ to enumerate time points before looping over space. i.e. $\mathbf{u} = \{ u_1^1,u_2^1,...,u_N^1, u_1^2,u_2^2,...,u_N^2, ...,u_N^k\}$, where subscript refers to time and superscript denotes space. This allows for more efficient assembly, because all data (coefficients in system of equation) for a specific element share memory locality, thus preventing cache misses. Moreover, this approach is compatible with existing frameworks for FEM, because problems with multiple DOFs are supported in existing FEM frameworks, so we can use standard procedures for system assembling or imposing boundary conditions. This has the additional advantage of reducing the matrix bandwidth. Table~\ref{bandwidth-table} enumerates the bandwidth for space-time formulation (2D mesh with $100 \times 100$ quad elements, linear basis function) using a Euler Backward formulation. As expected, the bandwidth linearly increases with increasing number of timesteps.

\begin{table}[bth!]
\caption{The bandwidth size for space-time formulation}
\label{bandwidth-table}
\begin{tabular}{|l|r|r|r|r|r|r|}
 \multicolumn{2}{c}{}  & \multicolumn{4}{c}{number of timesteps} \\
\cline{2-6}
 \multicolumn{1}{c|}{}   & sequential & 5 & 10 & 25 & 50 \\
\cline{1-6}
row size & 10201 & 51005 & 102010 & 255025 & 510050 \\ 
\hline
bandwidth & 205 & 1023 & 2043 & 5103 & 10203 \\
\hline
\end{tabular}
\end{table}

\begin{figure}[bth!]
\begin{center}
	\includegraphics[scale=0.75]{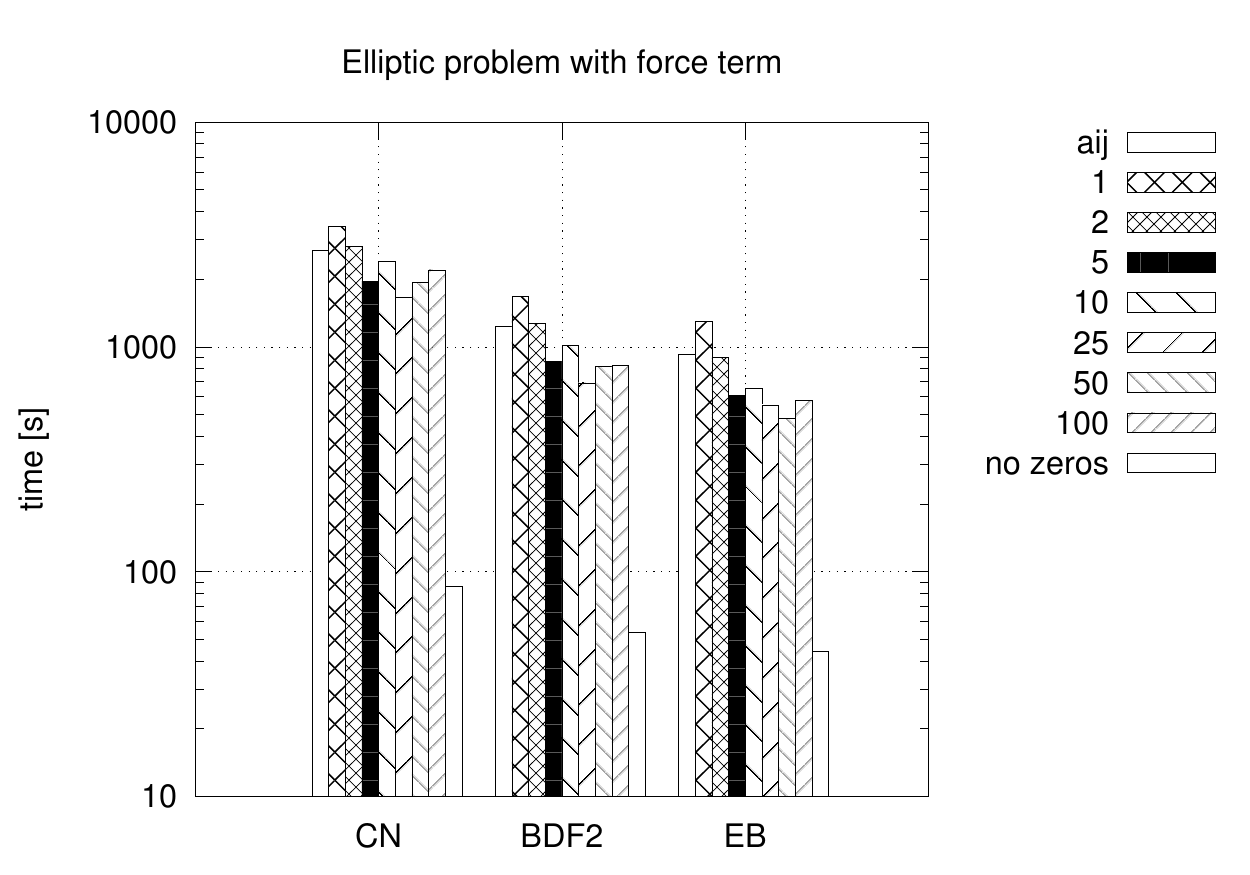}
	\caption{Computational time for different sizes of block size in block based matrix aij format \label{block-time-plot-stampede}. Investigated example was the 2D linear diffusion problem using 100x100 quad elements with one block of 100 timesteps. Numbers in legend describes block size, the 'aij' is sparse matrix without blocks and 'no zeroes' is special case with ignored zero entries in element matrix.}
\end{center}
\end{figure}

\subsection{Matrix access and storage}
\begin{figure}[bth!]
\begin{center}
	\includegraphics[scale=0.75]{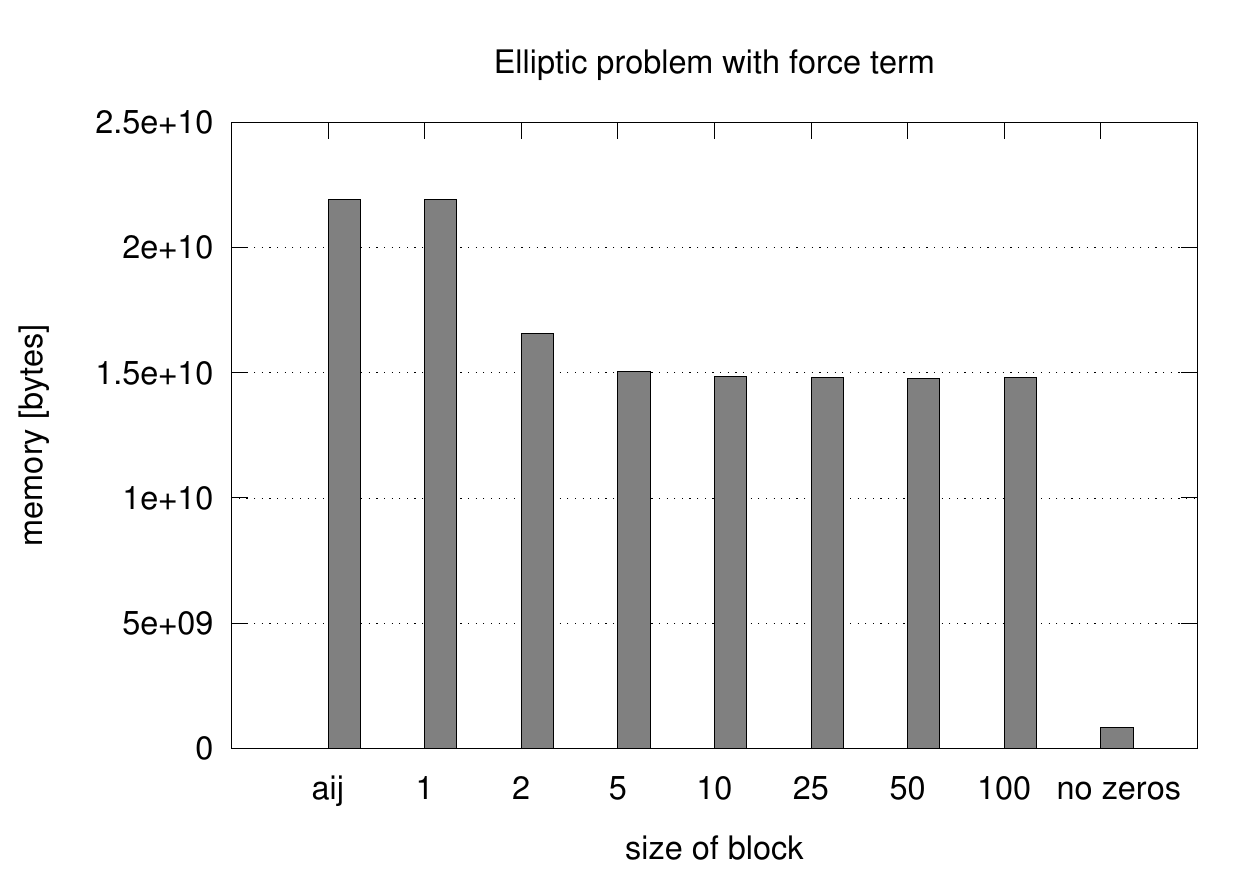}
	\caption{Memory requirements for different sizes of block size in block based matrix aij format and sparse element matrix\label{block-memory-stampede-with-zero-plot} On the example of elliptic problem on grid of 100x100 elements with one block of 100 timesteps}
\end{center}
\end{figure}

We explored the use of block structured matrix formats that are optimized for storing the global matrix, where the block size is a function of the number of timesteps. We use blocks of sizes ranging from $2 \times 2$ to $N \times N$ (where $N$ is the number of timesteps in a block, $B$). Figure \ref{block-time-plot-stampede} plots the time required to assemble and solve a linear diffusion problem using the space-time formulation (Backward Euler). We can see that different non-zero blocks sizes can give observable savings in computational time. Note also that each of these blocks is itself quite sparse (see  Figure.~\ref{compressed-element-matrix}). Careful identification of the non-zero pattern could potentially result in much larger savings. This is illustrated in the 'no-zero' column in Figure \ref{block-time-plot-stampede} where we avoid inserting zero elements to the global matrix.  The memory requirement for this case is also substantially minimized (see figure \ref{block-memory-stampede-with-zero-plot}). 

\subsection{Compressed element matrix}

\begin{figure}[bth!]
\begin{center}
	\includegraphics[scale=0.35]{linear_1_9EB_nodes.png}\hspace{1cm}
	\includegraphics[scale=0.35]{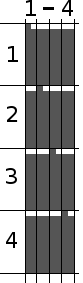}
	\caption{\label{compressed-element-matrix} View of full element matrix for Euler Backward (left) and compressed element matrix (right). Dark grey color means nonzero entries in matrix.}
\end{center}
\end{figure}

Figure \ref{compressed-element-matrix} shows the non zero patterns within an elemental stiffness matrix. The size of this matrix is $(nbf \times N) \times (nbf \times N)$, where $nbf$ is the number of basis functions (assuming 1 dof) and $N$ is the number of time steps in the block, $B$. Clearly, as N increases, the sparsity of the elemental stiffness matrix improves. This suggests using compressed formats for storing the elemental stiffness matrices. This issue becomes more important with increasing block size. 

\subsection{Scaling}

\begin{figure}[bth!]
\begin{center}
	\includegraphics[scale=0.65]{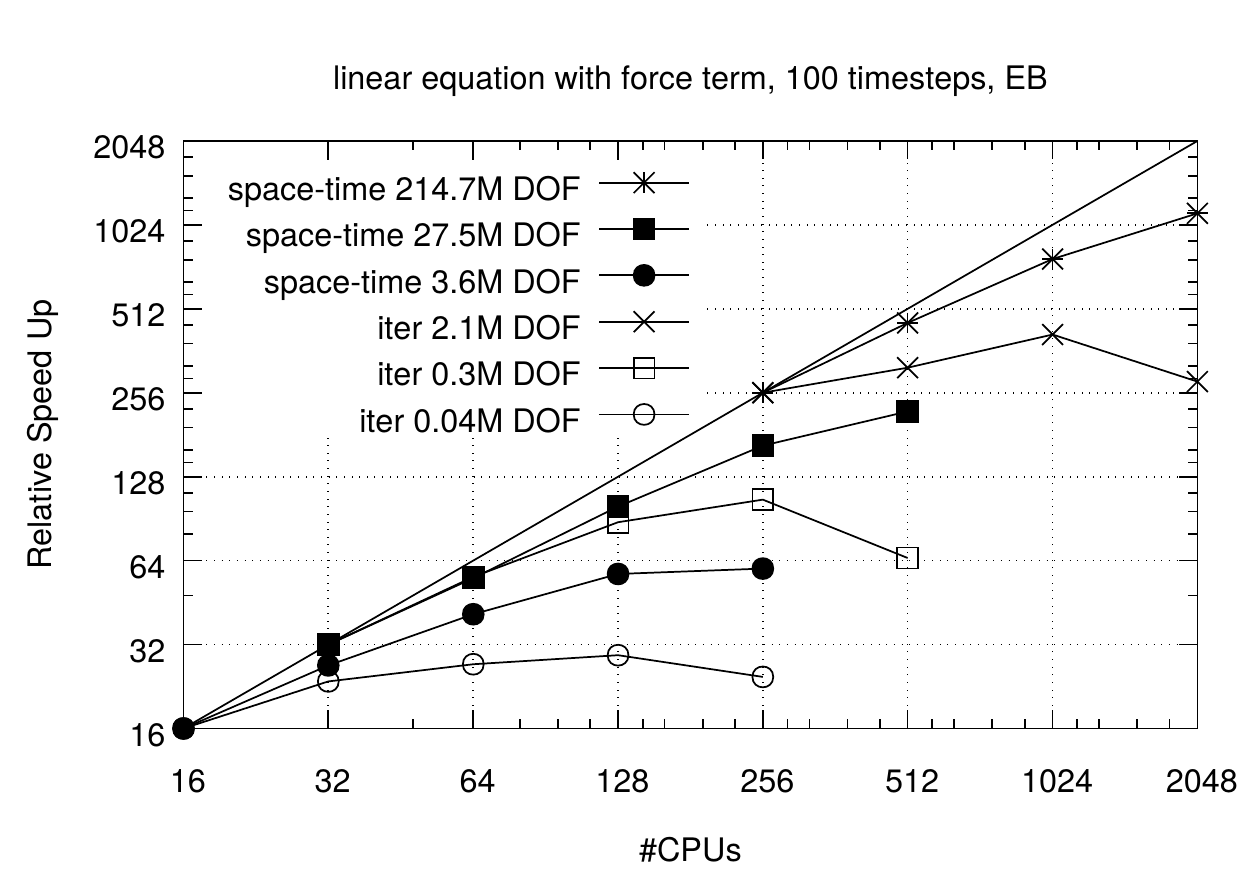}
	\includegraphics[scale=0.65]{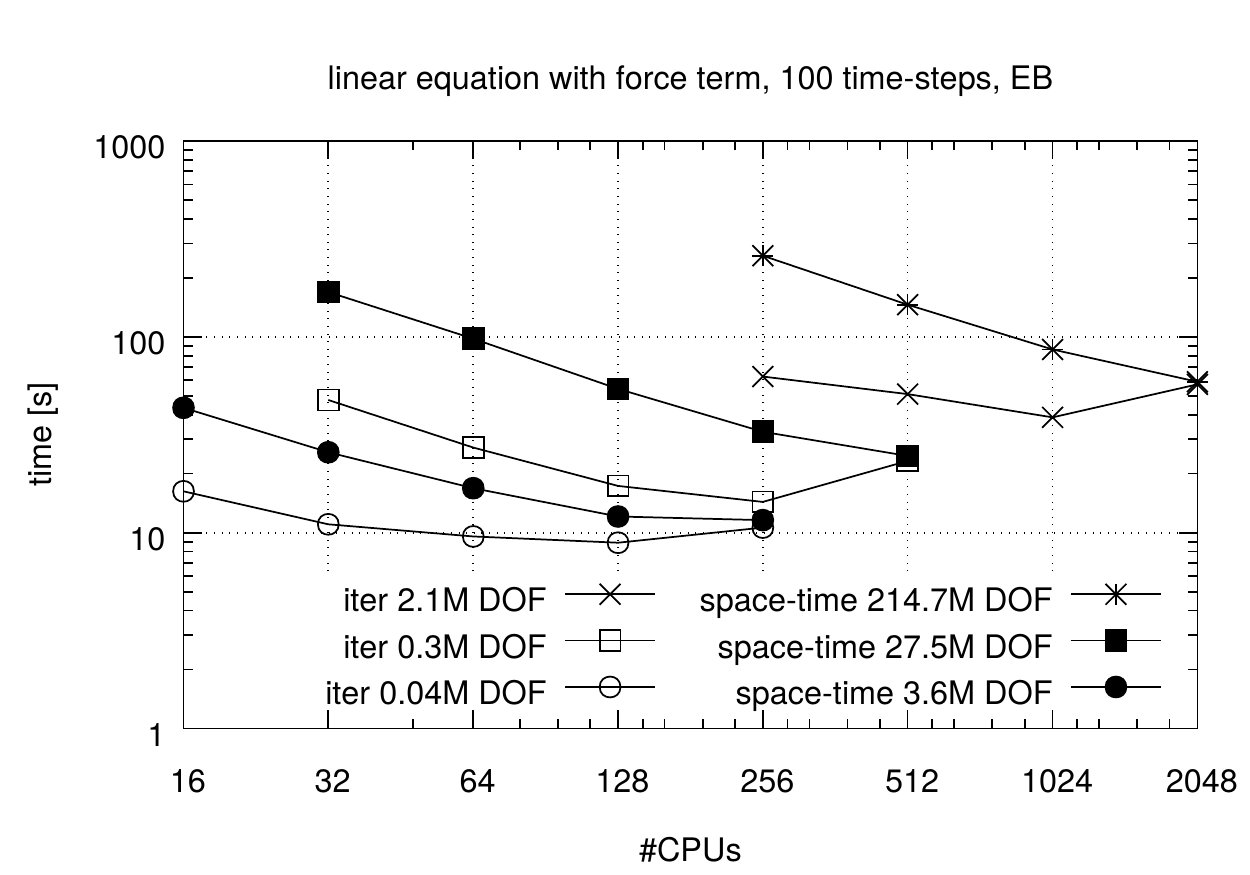}
	\caption{\label{scalability-stampede}Speed up on Stampede (top) and time requirements for space-time and iterative formulations  on the same machine (bottom)}
\end{center}
\end{figure}

We performed scaling studies on two machines. Preliminary scaling was performed on the TACC Stampede~\cite{stampede}. Stampede consists of 6400 compute nodes each equipped with two Intel E5-2680 8-core processors and 32 GB of memory. Each compute node has access to 250GB of local storage and Lustre filesystem with a write performance of 150GB/s. The nodes are connected with FDR InfiniBand network. Scaling results are shown in Figure \ref{scalability-stampede}. We also performed scaling on NCSA Blue Waters \cite{bw}. Blue Waters consists of 22,640 nodes, each consisting of two AMD 6276 "Interlagos" processors for total number of 362,240 computing cores. Each node has 64GB of memory, which gives total system memory equal to 1.476PB. Compute nodes have access to storage with Lustre filesystem that has size 26.4 PB and offers total bandwith of more than 1TB/s. 

\begin{figure}[bth!]
\begin{center}
	\includegraphics[scale=0.6]{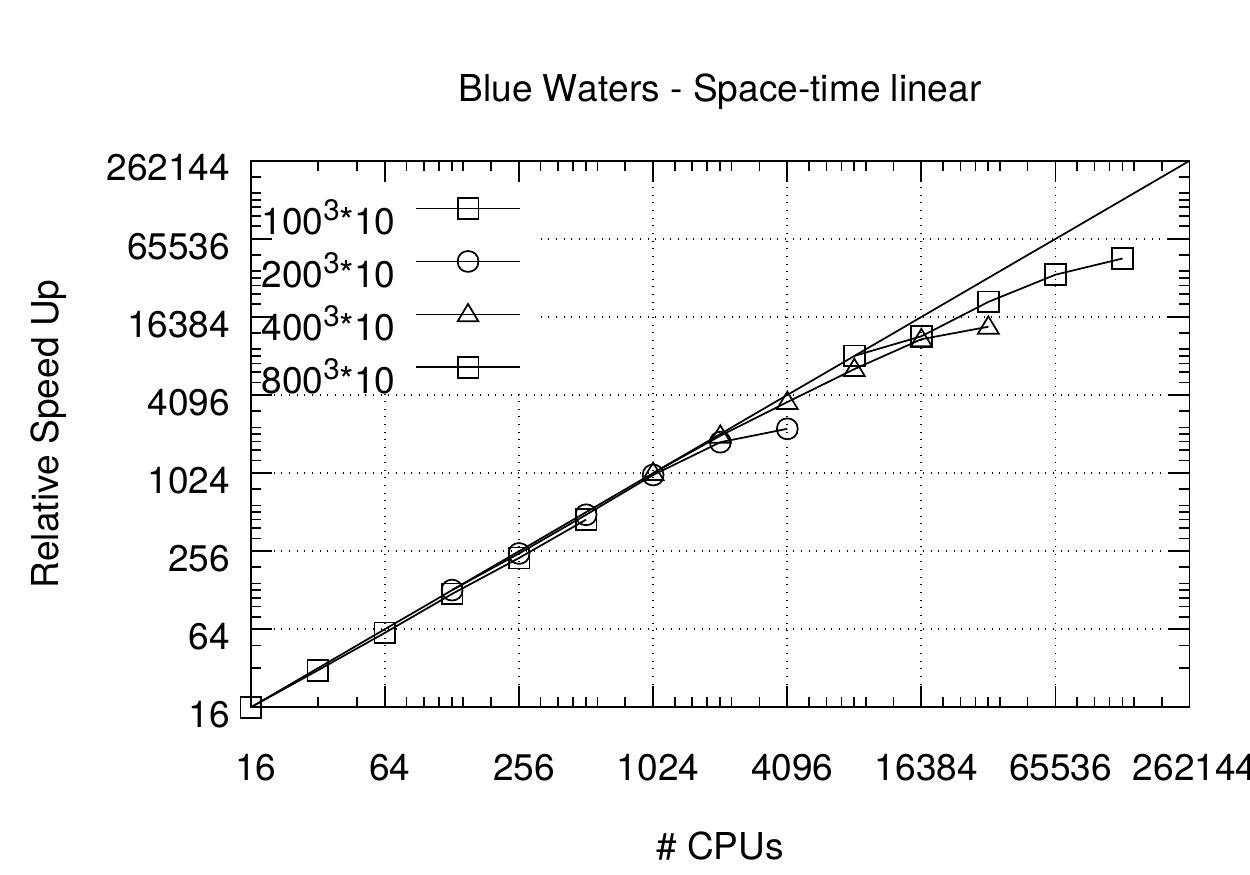}
	\caption{\label{linear-bw}Speed up on Blue Waters for linear time dependent diffusion}
\end{center}
\end{figure}

\begin{figure}[bth!]
\begin{center}
	\includegraphics[scale=0.6]{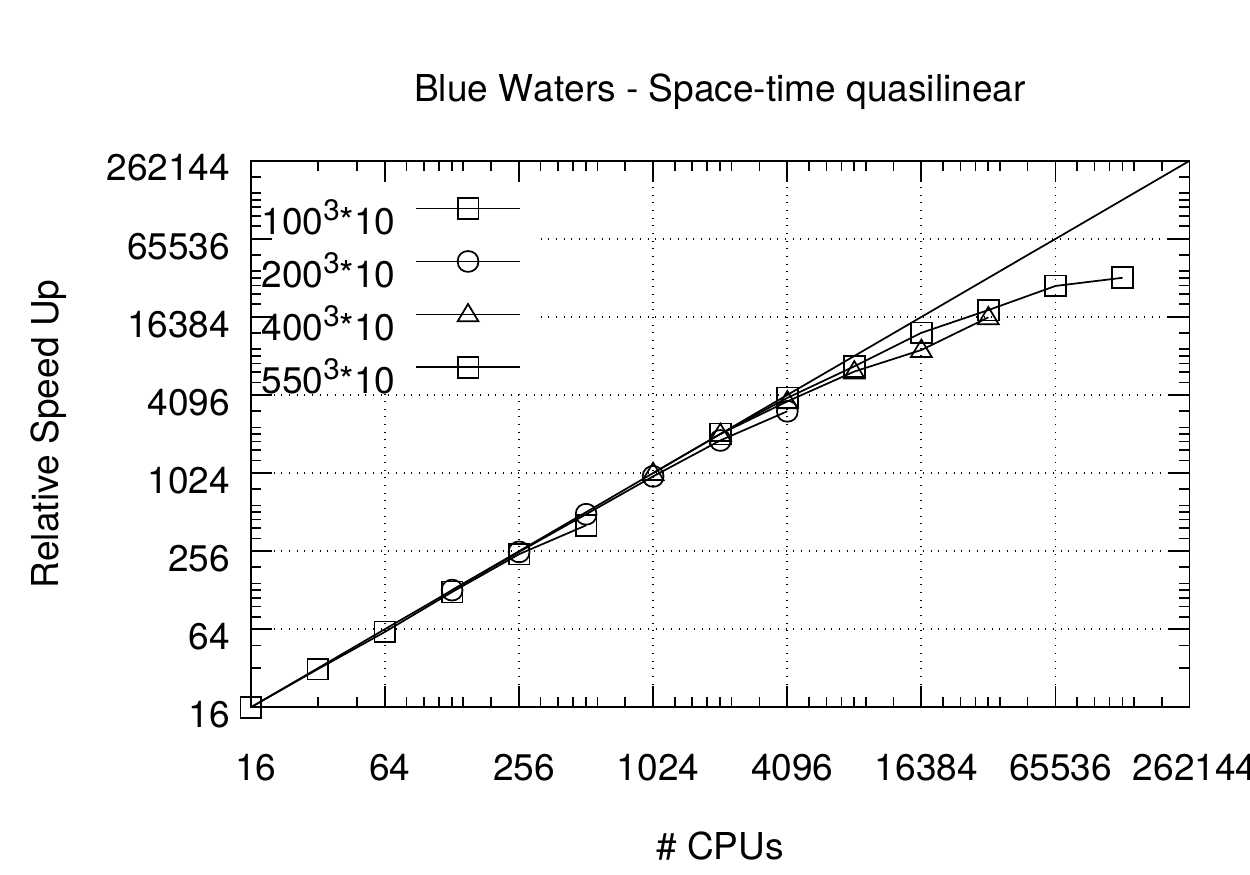}
	\caption{\label{nonlinear-bw}Speed up on Blue Waters for nonlinear time dependent diffusion (nonlinearity induced by diffusion coefficient depending on $u$)}
\end{center}
\end{figure}

\begin{figure}[bth!]
\begin{center}
	\includegraphics[scale=0.6]{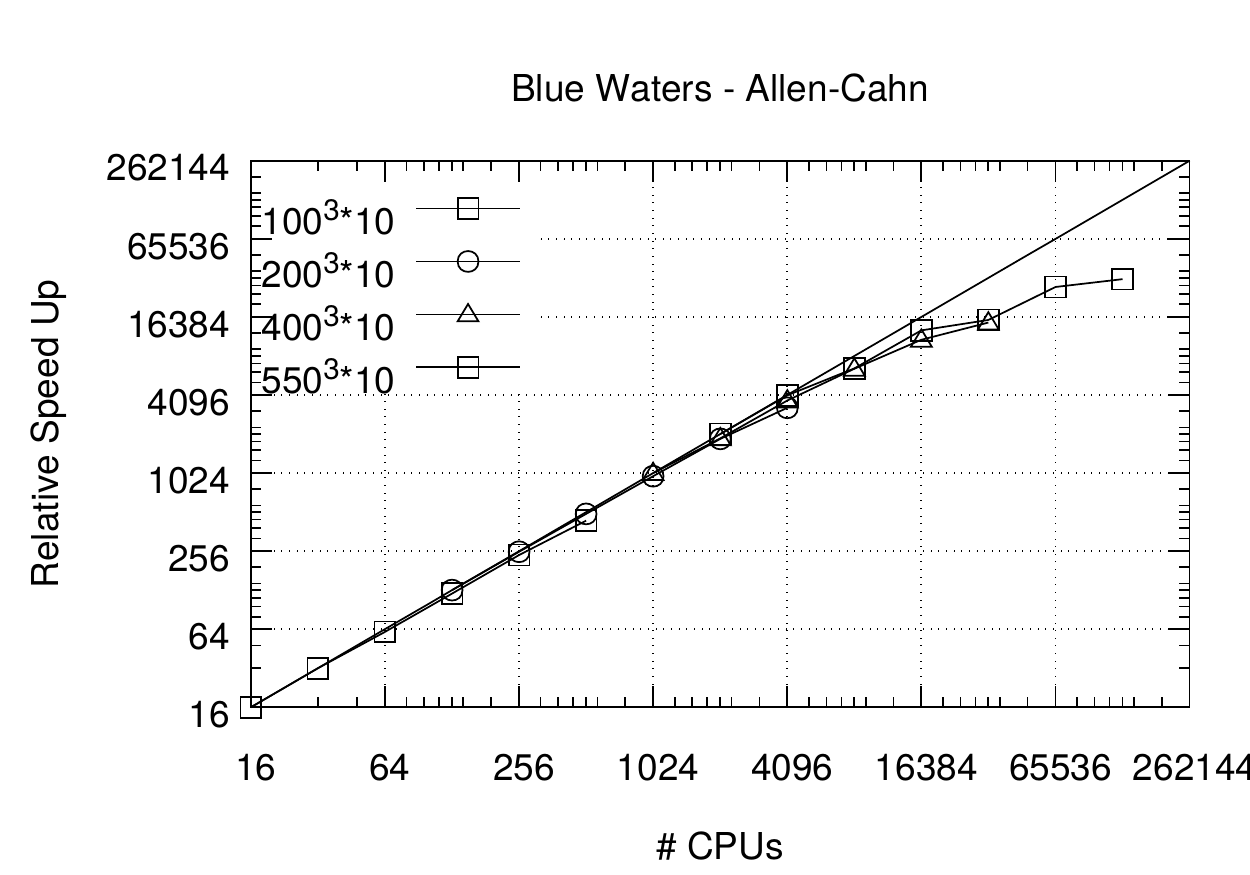}
	\caption{\label{ac-bw}Speed up on Blue Waters Supercomputer for Allen-Cahn problem}
\end{center}
\end{figure}

Figures \ref{linear-bw}, \ref{nonlinear-bw}, \ref{ac-bw} show scaling for the linear, nonlinear and Allen-Cahn equations respectively. We show good scaling upto 131,072 processors. The largest problem size used was $\sim 5.2$ billion degrees of freedom. This is a $800^3$ mesh with 10 timesteps per block.  
The results illustrate reasonably good scaling for the space-time approach, and Figure~\ref{scalability-stampede} suggests that for larger number of processors, the space-time approach performs better than the iterative approach. Specifically, beyond a threshold number of processors, the {\it total time to solve} a given problem is lower for the space time approach compared to the iterative approach. We anticipate that for more complex problems (complex geometries, remeshing, I/O) the space-time approach may perform even better.

\section{Numerical Examples}

We illustrate our adaptive parallel-in-space-time fra\-me\-work on linear and nonlinear diffusion equation and the Allen-Cahn equation.

\subsection{Problem A -- Linear Diffusion}
We first consider the linear case. We set $\kappa = 1$.  We use the method of manufactured solutions to construct the forcing term in \eqref{poisson} to ensure an analytical solution, $u$: 
\begin{equation}\label{linear_solution}
	u(x, y, z, t) = \exp\left( -\alpha(x, y, z, t) \right)
\end{equation}
where $\alpha(x,y,z,t)$ is equal to:
\begin{equation}
	\alpha(x,y,z,t) = \frac{(x-x_0(t))^2 + (y-y_0(t))^2 + (z-z_0(t))^2}{d^2}
\end{equation}
and
\begin{align*}
	x_0(t) &= a \cos(\omega t) + b \\
	y_0(t) &= a \sin(\omega t) + b \\
	z_0(t) &= b
\end{align*}
with $a = 0.2$, $b = 0.5$, $\omega = 2\pi$, $d=0.1$. Thus, $u$ is a rotating exponential hill centered at the midplane and rotating with a radius of 0.2 and an angular speed of $2 \pi$. This manufactured solution is constructed by the following forcing term
\begin{multline}
f(\mathbf{x}, t) = -\left( \frac{4((x-x_0)^2 + (y-y_0)^2 + (z-z_0)^2)}{d^4} - \frac{6}{d^2} \right. -\\
	\left. - 2 a \omega \left( \frac{\cos(t\omega)(y-y_0)-\sin(t\omega)(x-x_0)}{d^2} \right) \right)	
	\exp{\left(-\alpha(x,y,z,t)\right)}
\end{multline}
The equation is solved in a region of dimensions $[0:1]\times[0:1]\times[0:1]$. Every boundary face of the region has an essential boundary condition with prescribed value of $u$ equal to the value computed from \eqref{linear_solution}. We use a time step of 0.01 and solve for a block of 100 timesteps. 

\begin{figure}[bth!]
\begin{center}
    \includegraphics[width=0.35\textwidth]{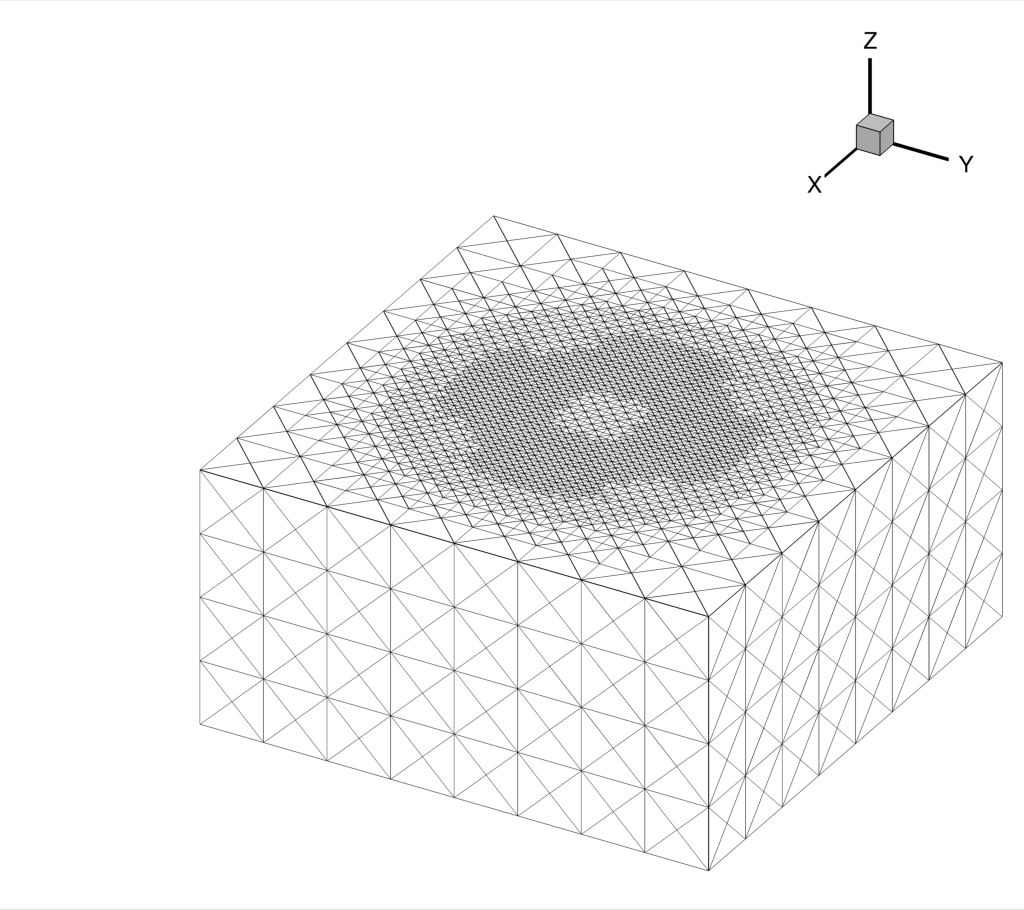}
	\includegraphics[width=0.35\textwidth]{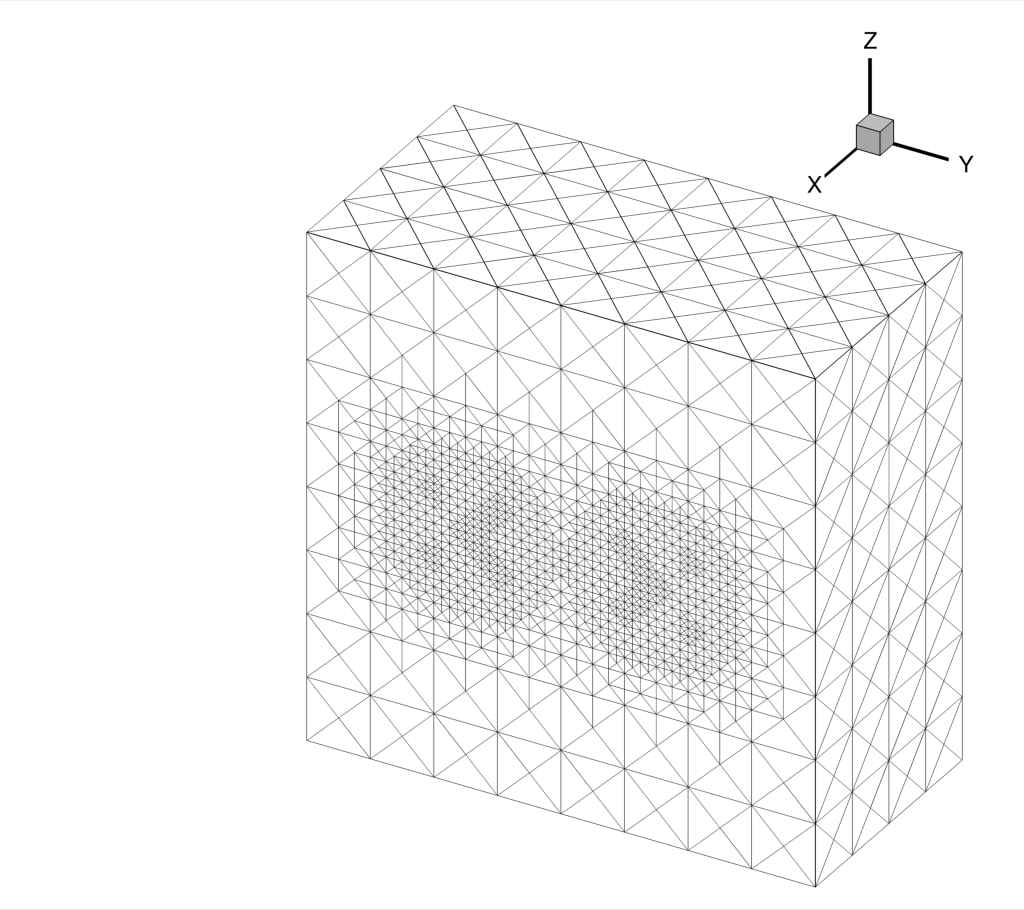}
    \caption{\label{A-mesh} Refined mesh: view from top (left) and front (right).}
\end{center}
\end{figure}

Figure~\ref{A-mesh} shows a cutoff of the refined spatial mesh after 20 refinement iterations. We remind the reader that each iteration consists of solving the space-time problem, constructing the time-averaged elemental refinement indicators, equation~\eqref{eqn:error_indicator1}, and then refining the 3D mesh according to the indicators. Given that this is a moving source problem, it is clearly seen that there is refinement in regions where the source has passed through. 

\begin{figure}[bth!]
\begin{center}
    \includegraphics[scale=0.6]{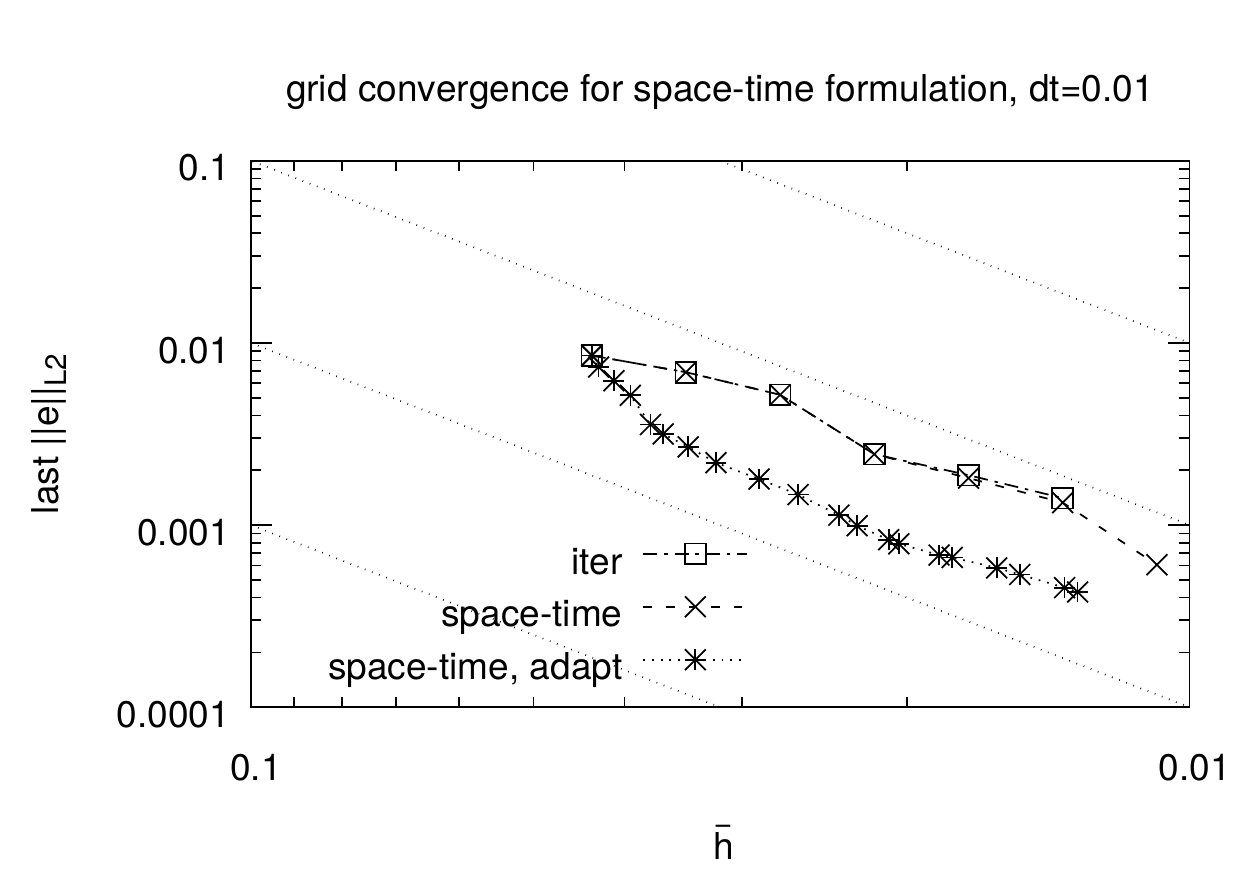}
    \caption{\label{A-space-error} Grid convergence for linear heat equation. Error is $\|u - u_h\|_2$ from last time-step versus average element size $\bar{\text{h}}$.}
\end{center}
\end{figure}

\begin{figure}[bth!]
\begin{center}
    \includegraphics[scale=0.6]{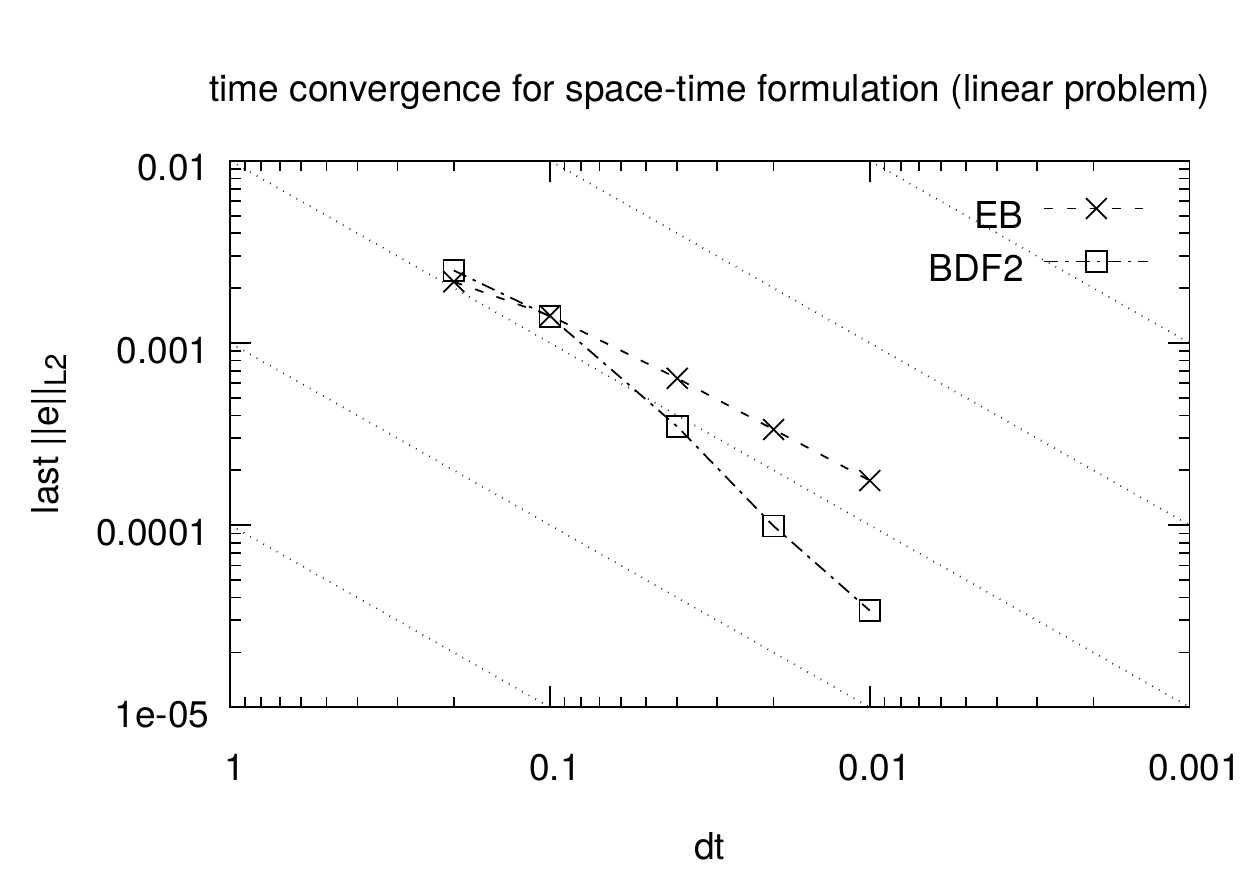}
    \caption{\label{A-time-error} Time convergence for linear heat equation. Error is $\|u - u_h\|_2$ from last time-step versus size of time-step.}
\end{center}
\end{figure}

We compare spatial convergence rates for three implementations: (a) sequential time stepping with no spatial adaptivity, (b) space-time implementation with no spatial adaptivity, and (c) space-time implementation with spatial adaptivity. We plot convergence in Figure~\ref{A-space-error}, where error is  $\|u - u_h\|_2$. The first two implementations (obviously) overlap, with the adaptive mesh implementation showing a reduced error. All three curves show a slope of 2, which is to be expected. The Figure~\ref{A-time-error} shows time-step convergence, with expected slopes of 1 and 2 for Backward Euler and Backward Difference Formulae, respectively.

\subsection{Problem B -- Nonlinear Diffusion}
For the nonlinear case, we set the coefficient $\kappa(u)=1+10u^2$ and we choose an exact solution such that $ |u| \leq 1$, thus bounding $\kappa$. As before, we choose our analytical solution to be given by~\eqref{linear_solution}. The forcing term, $f$, is consequently:
\begin{multline}
f(\mathbf{x}, t) = \\
	\frac{80u}{d^4}\left( (x-x_0)^2 + (y-y_0)^2 + (z-z_0)^2 \right)
	\exp{\left(-2\alpha(x, y, z, t)\right)} + \\
	+ (1+10u^2)\left( \frac{4((x-x_0)^2 + (y-y_0)^2 + (z-z_0)^2)}{d^4} - \frac{6}{d^2} \right) 
	\exp{\left(-\alpha(x, y, z, t)\right)} - \\
	- 2 a \omega \frac{\cos(t\omega)(y-y_0)-\sin(t\omega)(x-x_0)}{d^2}
	\exp{\left(-\alpha(x, y, z, t)\right)}
\end{multline}

\begin{figure}[bth!]
\begin{center}
	\begin{subfigure}[b]{0.4\textwidth}
	    \includegraphics[width=\textwidth]{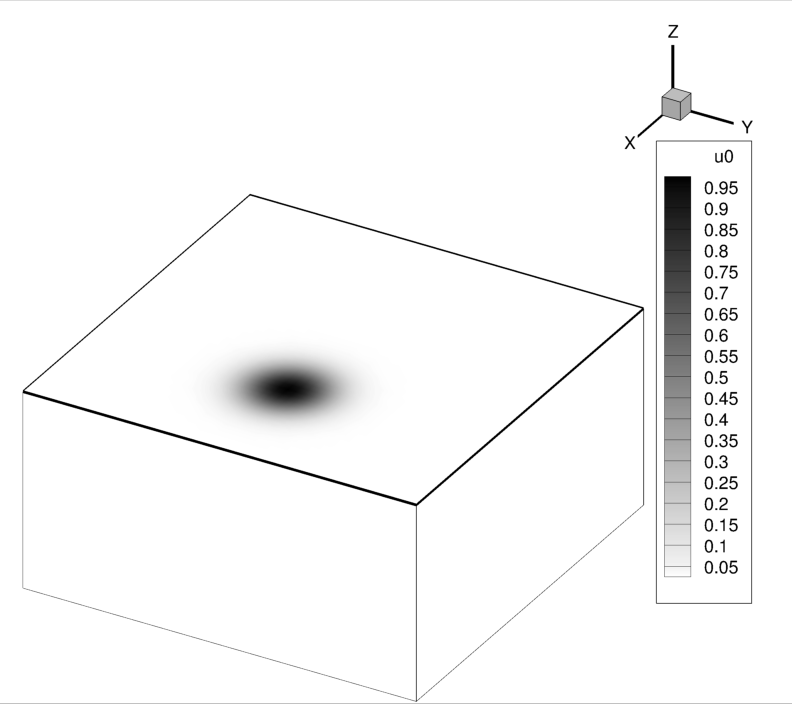}
	    \caption{\label{rot-0}t=0s}
	\end{subfigure}
	\begin{subfigure}[b]{0.4\textwidth}
	    \includegraphics[width=\textwidth]{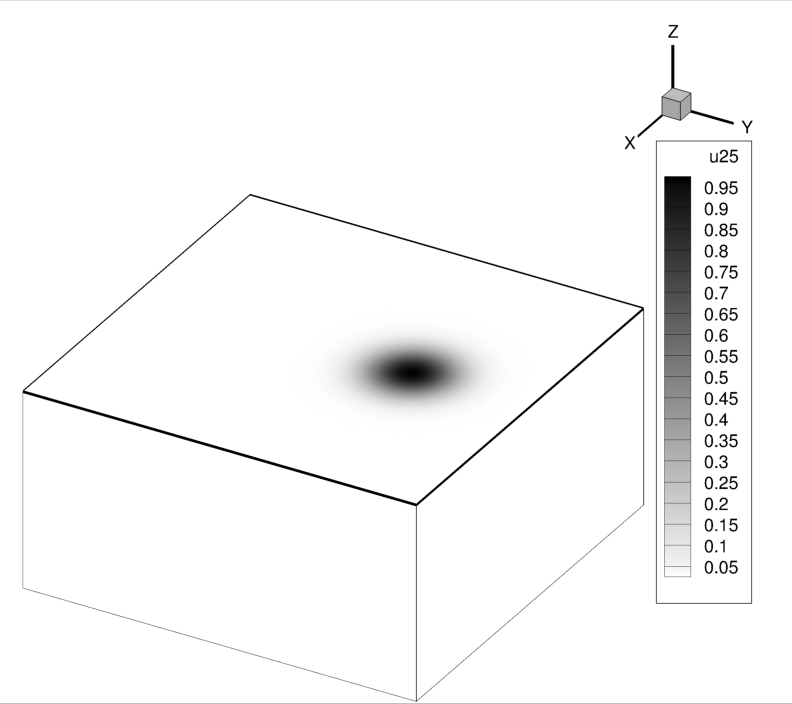}
	    \caption{\label{rot-25}t=0.25s}
	\end{subfigure}
	\begin{subfigure}[b]{0.4\textwidth}
	    \includegraphics[width=\textwidth]{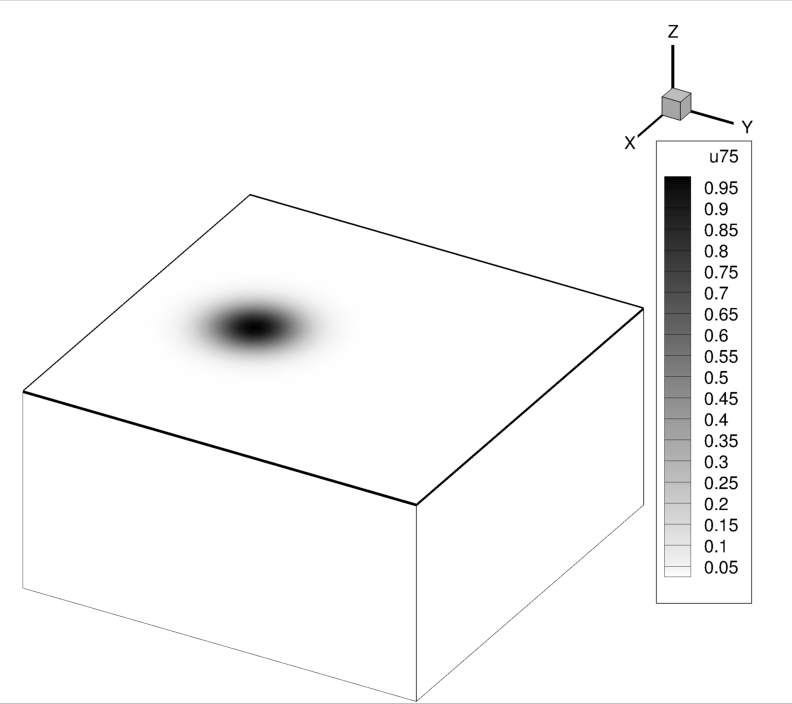}
	    \caption{\label{rot-75}t=0.75s}
	\end{subfigure}
	\begin{subfigure}[b]{0.4\textwidth}
	    \includegraphics[width=\textwidth]{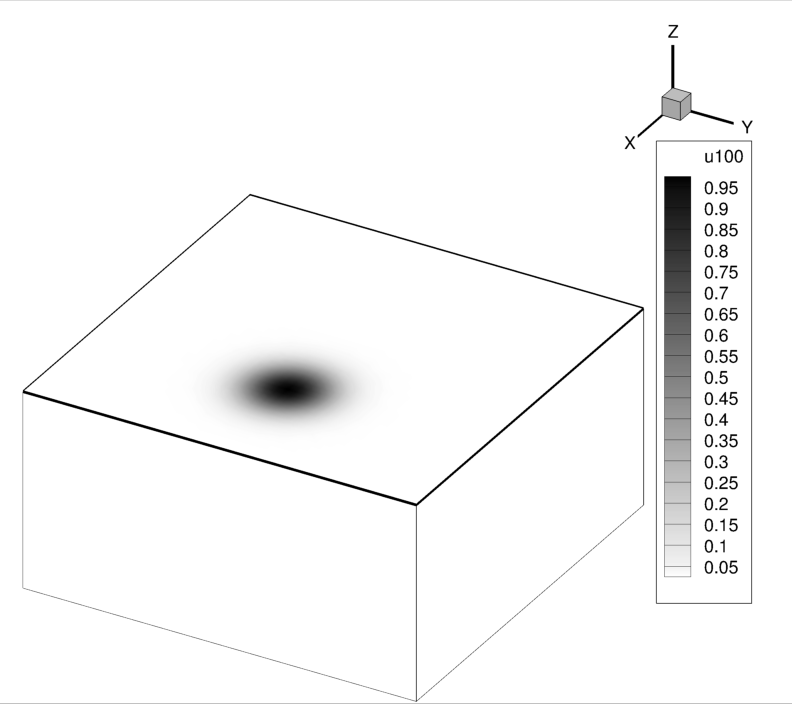}
	    \caption{\label{rot-100}t=1s}
	\end{subfigure}
	\caption{\label{A-u} Results of computations where upper half of computational domain is removed. 
    	Values of $u$ change in time from 0s (a) to 1s (d)}
\end{center}
\end{figure}

\begin{figure}[bth!]
\begin{center}
    \includegraphics[scale=0.6]{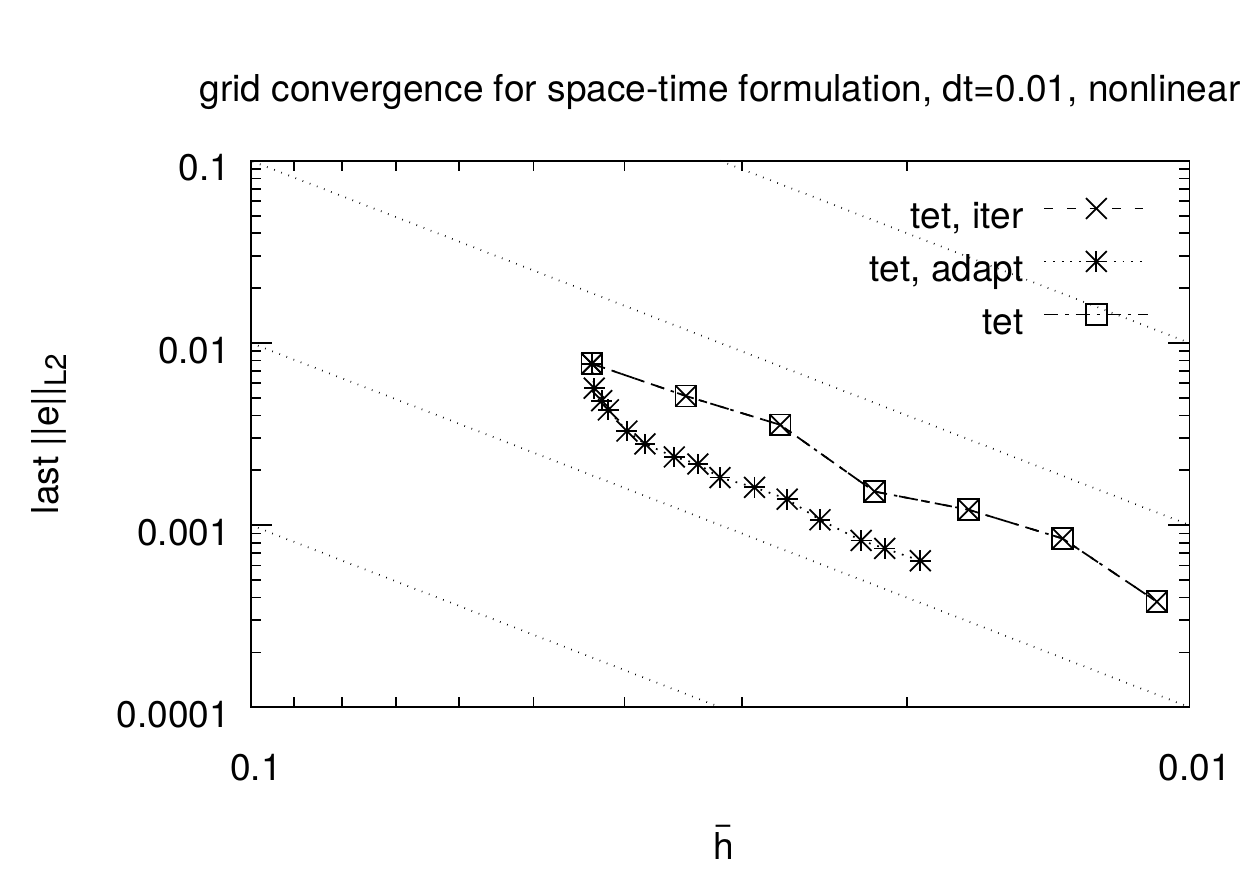}
    \caption{\label{B-error} Grid convergence for nonlinear heat equation. Error is $\|u - u_h\|_2$ from last time-step versus average element size $\bar{\text{h}}$.}
\end{center}
\end{figure}

Figure.~\ref{A-u} plots several time snapshots of the moving non-linear source problem, with the solution accurately tracking the moving source. In Figure.~\ref{B-error}, we plot spatial convergence for three implementations:  (a) sequential time stepping with no spatial adaptivity, (b) space-time implementation with no spatial adaptivity, and (c) space-time implementation with spatial adaptivity (plotted with the average element size). Convergence rates follow along expected lines, with a slope of 2. % and order 1 in case of timestep convergence (figure \ref{B-timestep}).

\subsection{Problem C -- Allen-Cahn}
\begin{figure}[bth!]
\begin{center}
	\begin{subfigure}[b]{0.35\textwidth}
	    \includegraphics[width=\textwidth]{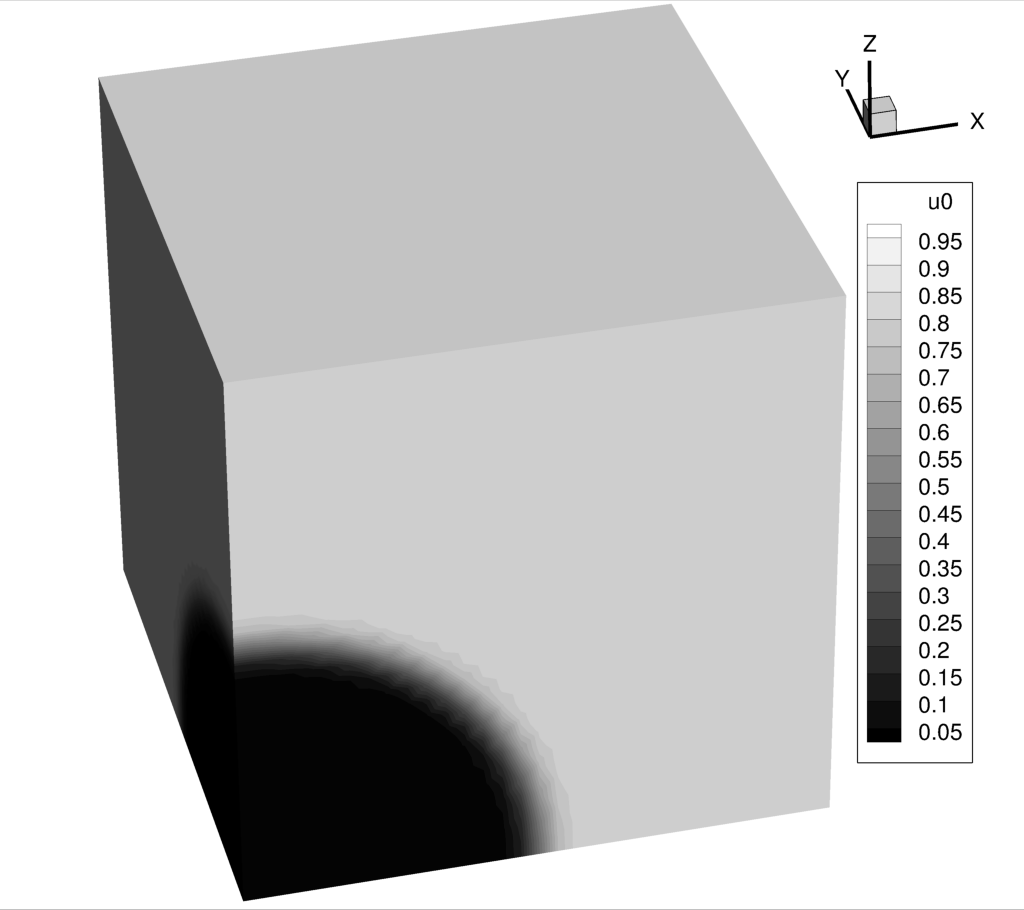}
	    \caption{\label{rot-0AC}t=0s}
	\end{subfigure}
	\begin{subfigure}[b]{0.35\textwidth}
	    \includegraphics[width=\textwidth]{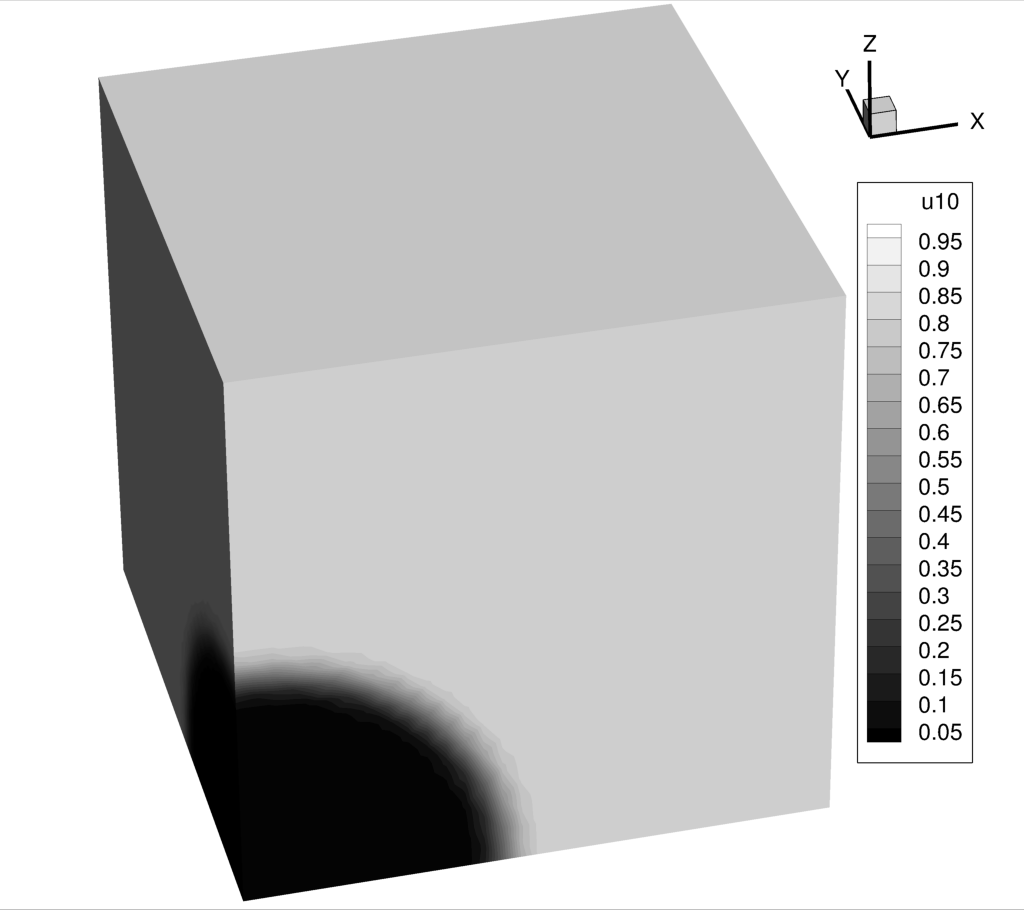}
	    \caption{\label{rot-25AC}t=1s}
	\end{subfigure}
	\begin{subfigure}[b]{0.35\textwidth}
	    \includegraphics[width=\textwidth]{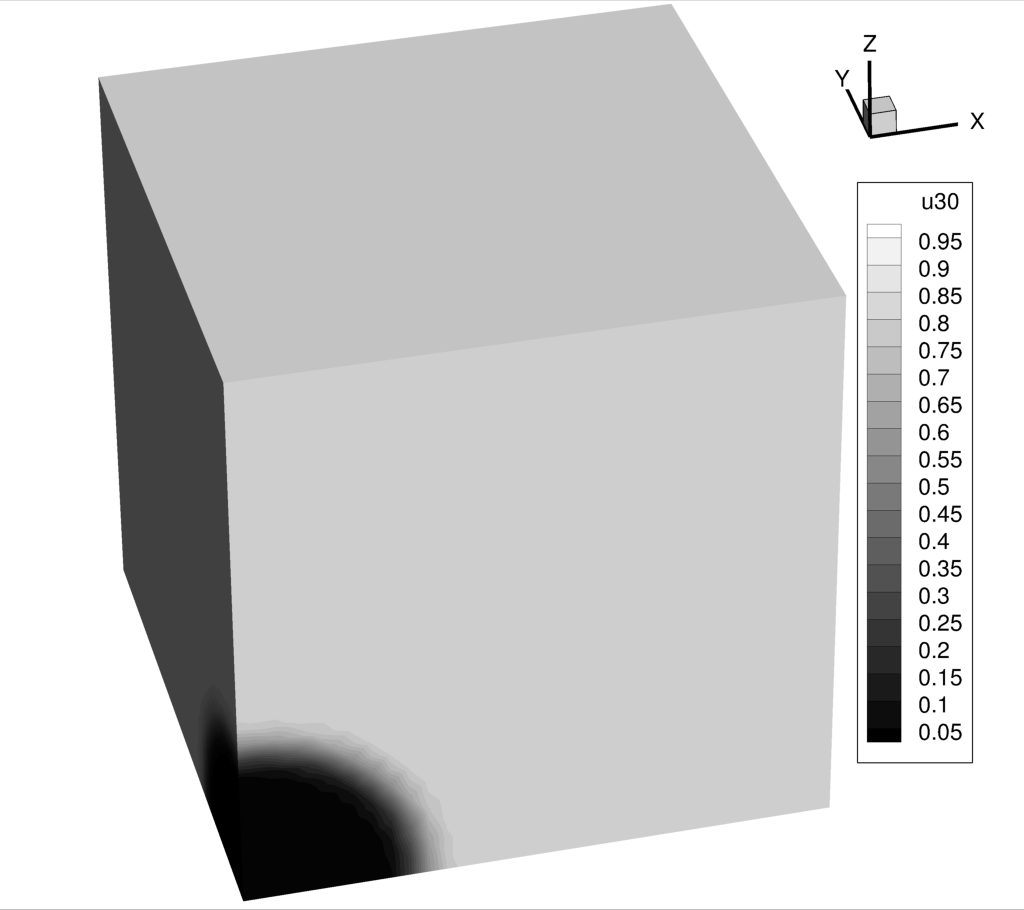}
	    \caption{\label{rot-75AC}t=3s}
	\end{subfigure}
	\begin{subfigure}[b]{0.35\textwidth}
	    \includegraphics[width=\textwidth]{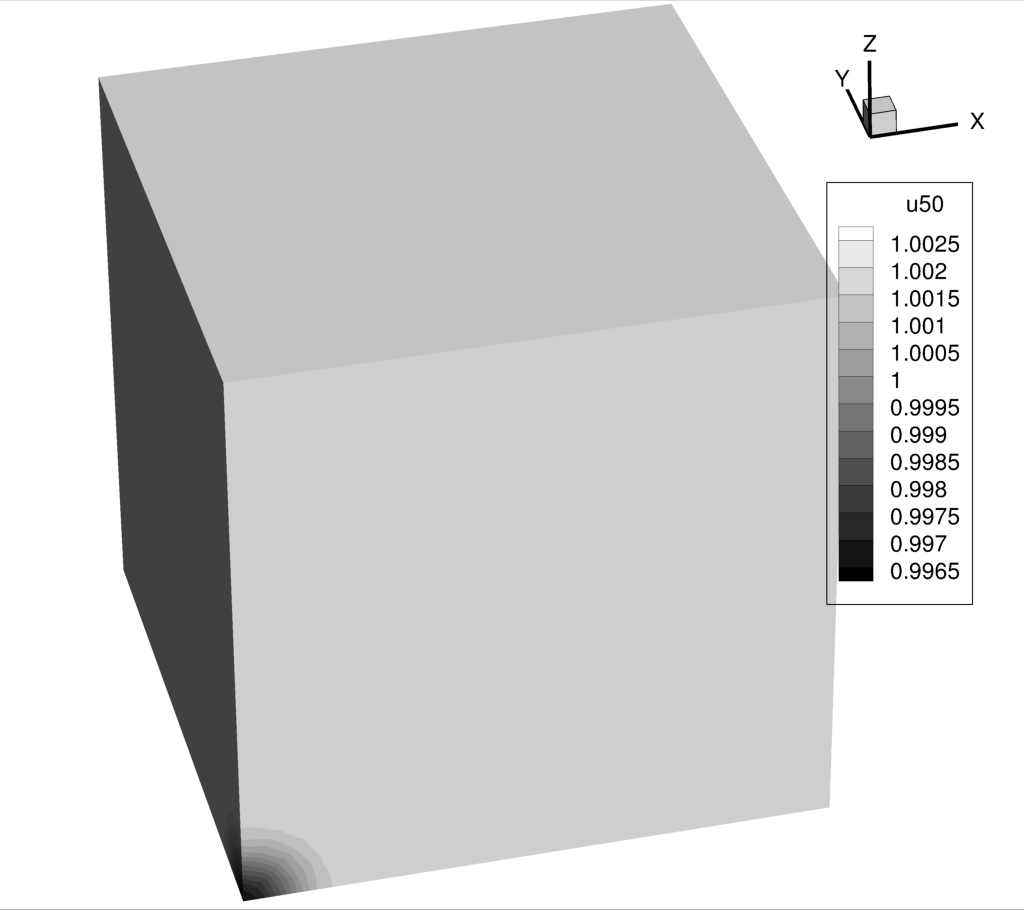}
	    \caption{\label{rot-100AC}t=5s}
	\end{subfigure}
	\caption{\label{AC-u} Results of computations for the 3D Allen-Cahn problem. Values of $u$ for initial condition (a), after 1s (b), 3s (c) and 5s (d)}
\end{center}
\end{figure}

In this final example we solve the modified Allen-Cahn problem. The key physics (i.e. phase change) which is described by this non-linear equation essentially occurs in a highly localized region of the domain (on a surface of co-dimension 1). Adaptive refinement (and coarsening) has been a very effective approach to accurately resolve this localized region. The equation is given as $\frac{\partial u}{\partial t} = -D \left( f(u) - C_n^2 \nabla^2 u \right)$, where
\begin{equation}
	f(u) = 2 A u (1 - 3 u + 2 u^2) - k
\end{equation}
and 	$D=1$, $C_n=0.1$, $A=16$, $k=0.1$. The initial conditions are
\begin{equation}
u(x) = 0.5 + 0.5 \tanh \left( \frac{r-0.5}{\sqrt{\frac{2}{A}}C_n} \right), \qquad r = \sqrt{x^2 + y^2 + z^2}
\end{equation}
with zero flux conditions on all boundaries. This represents an initial solid of radius, $r=0.5$, that is melting. Using symmetry arguments, we consider a single octant of the space ($[0:1]\times[0:1]\times[0:1]$). We consider a time step of 0.02 and a block size of 50 time-steps. 
Figure~\ref{AC-u} shows snapshots of the melting sphere at various time points.

\begin{figure}[bth!]
\begin{center}
	\includegraphics[width=0.3\textwidth]{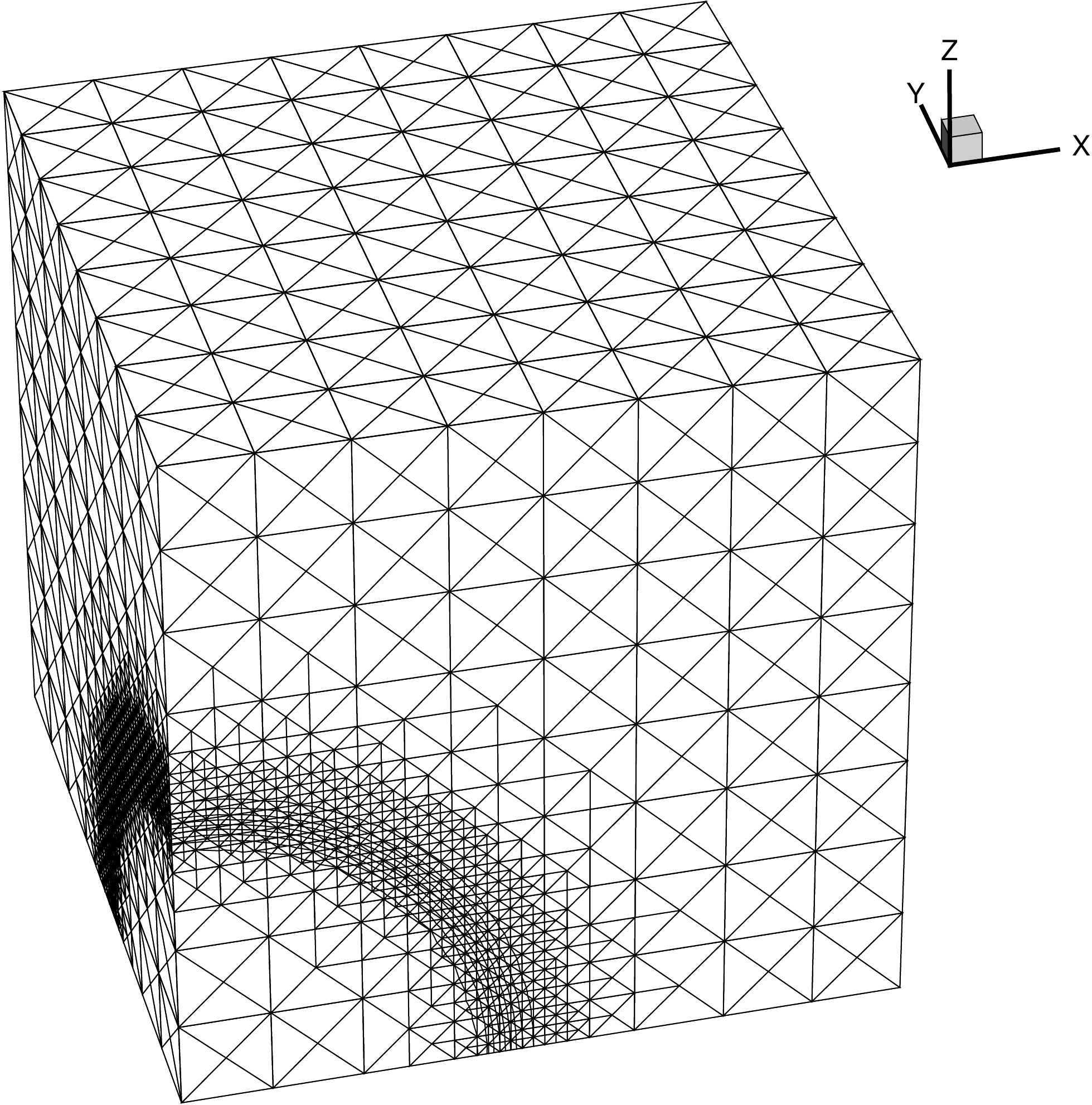}
    \caption{\label{ac_3d_mesh} Finite element mesh after 20 refinement iterations with superimposed solution at time $1s$ at the dense part of mesh}
\end{center}
\end{figure}

Figure~\ref{ac_3d_mesh} illustrates adaptive mesh refinement across the time block. The figure shows the refined mesh for the first block of time ($t=0$ to $t=1$) Note that refinement is localized along the solidification front {\it within the time block}. To capture this thin interface using a uniform mesh would have required 589~824 elements, instead of the 137~776 elements that were used here.

\section{Conclusion}
We present formulation, implementation details and representative examples of a parallel-in-space-time based adaptive methodology for the solution of  (linear and) non-linear time dependent problems. The basic concept is to solve for large blocks of space-time unknowns instead of marching sequentially in time. The methodology is a combination of a computationally efficient implementation of a parallel-in-space-time finite element solver coupled with {\it a posteriori} space-time error estimates and a parallel mesh generator. We illustrate how this implementation is especially tailored for massively parallel computations. We show good scaling behavior up to 150,000 processors on the Blue Waters machine. This methodology enables scaling on next generation multi-core machines by simultaneously solving for large number of time-steps, and reduces computational overhead by refining spatial blocks that can track localized features. This methodology also opens up the possibility of efficiently incorporating adjoint equations for error estimators and inverse design problems, since blocks of space-time are simultaneously solved and stored in memory.  Our future work is focused on extending the space-time framework to utilizing finite element basis functions in time (which enables formal derivation of space-time a posteriori error estimates), and subsequently implementing 4D finite elements to enable simultaneous space and time adaptivity.

\bibliographystyle{amsplain}
\bibliography{references}

\providecommand{\bysame}{\leavevmode\hbox to3em{\hrulefill}\thinspace}
\providecommand{\MR}{\relax\ifhmode\unskip\space\fi MR }
\providecommand{\MRhref}[2]{%
  \href{http://www.ams.org/mathscinet-getitem?mr=#1}{#2}
}
\providecommand{\href}[2]{#2}
\begin{thebibliography}{10}

\bibitem{bw}
\emph{Blue waters user portal \textbar{} system summary}.

\bibitem{phg}
\emph{Phg (parallel hierarchical grid)}.

\bibitem{stampede}
\emph{Texas advanced computing center - stampede technical details}.

\bibitem{ainsworth_oden}
Mark Ainsworth and J.~Tinsley Oden, \emph{A posteriori error estimation in
  finite element analysis}, John Wiley \& Sons, New York, 2000.

\bibitem{petsc}
Satish Balay, William~D. Gropp, Lois~Curfman McInnes, and Barry~F. Smith,
  \emph{Efficient management of parallelism in object oriented numerical
  software libraries}, Modern Software Tools in Scientific Computing (E.~Arge,
  A.~M. Bruaset, and H.~P. Langtangen, eds.), Birkh{\"{a}}user Press, 1997,
  pp.~163--202.

\bibitem{behr}
Marek Behr, \emph{Simplex space-time meshes in finite element simulations},
  Int. J. Numer. Meth. Fluids \textbf{57} (2008), no.~9, 1421--1434.

\bibitem{burrage}
Kevin Burrage, \emph{10. parallel methods for systems of ordinary differential
  equations}, Applications on Advanced Architecture Computers, 1996,
  pp.~101--120.

\bibitem{butcher}
John~C. Butcher, \emph{Numerical methods for ordinary differential equations},
  John Wiley \& Sons, West Sussex, 2008.

\bibitem{carey_estep}
V.~Carey, D.~Estep, A.~Johansson, M.~Larson, and S.~Tavener, \emph{Blockwise
  adaptivity for time dependent problems based on coarse scale adjoint
  solutions}, SIAM J. Sci. Comput. \textbf{32} (2010), no.~4, 2121--2145.

\bibitem{simsek_zee}
G.~\c{S}im\c{s}ek, X.~Wu, K.G. van~der Zee, and E.H. van Brummelen,
  \emph{Duality-based two-level error estimation for time-dependent pdes:
  Application to linear and nonlinear parabolic equations}, Computer Meth.
  Appl. Mech. Engrg. \textbf{288} (2015), 83--109.

\bibitem{debusschere}
Bert~J. Debusschere, Habib~N. Najm, Philippe~P. Pébay, Omar~M. Knio, Roger~G.
  Ghanem, and Olivier P.~Le Maı⁁tre, \emph{Numerical challenges in the use
  of polynomial chaos representations for stochastic processes}, SIAM J. Sci.
  Comput. \textbf{26} (2004), no.~2, 698--719.

\bibitem{eriksson_johnson}
K.~Eriksson, C.~Johnson, and A.~Logg, \emph{Adaptive computational methods for
  parabolic problems}, Encyclopedia of Computational Mechanics, in:
  Fundamentals (E.~Stein, R.~de~Borst, and T.J.R. Hughes, eds.), John Wiley \&
  Sons, Ltd., 2004, pp.~675--702.

\bibitem{gomez_zee}
Hector Gomez and Kristoffer~G. van~der Zee, \emph{Computational phase-field
  modeling}, Encyclopedia of Computational Mechanics, Second Edition, John
  Wiley \& Sons, Ltd., (In Press).

\bibitem{siam2013}
Jeffrey Hittinger, Sven Leyffer, and Jack Dongarra, \emph{Models and algorithms
  for exascale computing pose challenges for applied mathematicians}, SIAM
  News, 2013.

\bibitem{hughes_hulbert}
Thomas J.~R. Hughes and G.~M. Hulbert, \emph{Space-time finite element methods
  for elastodynamics: Formulations and error estimates}, Comput. Methods Appl.
  Mech. Eng. \textbf{66} (1988), no.~3, 339--363.

\bibitem{hughes_steward}
Thomas~J.R. Hughes and James~R. Stewart, \emph{A space-time formulation for
  multiscale phenomena}, J. Comput. Appl. Math. \textbf{74} (1996), no.~1,
  217--229.

\bibitem{parareal2001}
J.~Lions, Yvon Maday, and Gabriel Turinici, \emph{A''parareal''in time
  discretization of pde's}, C. R. Acad. Sci. Series I Mathematics \textbf{332}
  (2001), no.~7, 661--668.

\bibitem{lowrie1996}
Robert~B. Lowrie, Philip~L. Roe, and Bram van Leer, \emph{Space-time methods
  for hyperbolic conservation laws}, Barriers and Challenges in Computational
  Fluid Dynamics (V.~Venkatakrishnan, Manuel~D. Salas, and Sukumar~R.
  Chakravarthy, eds.), Springer Netherlands, Dordrecht, 1998, pp.~79--98.

\bibitem{mani2011}
Karthik Mani and Dimitri Mavriplis, \emph{Efficient solutions of the euler
  equations in a time-adaptive space-time framework}, 49th AIAA Aerospace
  Sciences Meeting including the New Horizons Forum and Aerospace Exposition,
  American Institute of Aeronautics and Astronautics, 2011.

\bibitem{narayan}
Akil Narayan and Dongbin Xiu, \emph{Stochastic collocation methods on
  unstructured grids in high dimensions via interpolation}, SIAM J. Sci.
  Comput. \textbf{34} (2012), no.~3, A1729--A1752.

\bibitem{potanza_reddy}
J.~P. Pontaza and J.~N. Reddy, \emph{Space-time coupled spectral/hp
  least-squares finite element formulation for the incompressible navier-stokes
  equations}, J. Comput. Phys. \textbf{197} (2004), no.~2, 418--459.

\bibitem{rendall2012}
Thomas C.~S. Rendall, Christian~B. Allen, and Edward D.~C. Power,
  \emph{Conservative unsteady aerodynamic simulation of arbitrary boundary
  motion using structured and unstructured meshes in time}, Int. J. Numer.
  Meth. Fluids \textbf{70} (2012), no.~12, 1518--1542.

\bibitem{soize}
Christian Soize and Roger Ghanem, \emph{Physical systems with random
  uncertainties: Chaos representations with arbitrary probability measure},
  SIAM J. Sci. Comput. \textbf{26} (2004), no.~2, 395--410.

\bibitem{tezduyar}
Tayfun~E. Tezduyar, Sunil Sathe, Ryan Keedy, and Keith Stein,
  \emph{Space–time finite element techniques for computation of
  fluid–structure interactions}, Computer Meth. Appl. Mech. Engrg.
  \textbf{195} (2006), no.~17--18, 2002--2027.

\bibitem{verfurth1996}
R\"{u}diger Verf\"{u}rth, \emph{A review of a posteriori error estimation and
  adaptive mesh-refinement techniques}, Wiley-Teubner, New York and Stuttgart,
  1996.

\bibitem{verfurth2013}
\bysame, \emph{A posteriori error estimation techniques for finite element
  methods}, Oxford University Press, Oxford, 2008.

\bibitem{wan}
Xiaoliang Wan and George~Em Karniadakis, \emph{Multi-element generalized
  polynomial chaos for arbitrary probability measures}, SIAM J. Sci. Comput.
  \textbf{28} (2006), no.~3, 901--928.

\bibitem{wang2015}
Luming Wang and Per-Olof Persson, \emph{A high-order discontinuous galerkin
  method with unstructured space–time meshes for two-dimensional compressible
  flows on domains with large deformations}, Comput Fluids \textbf{118} (2015),
  53--68.

\bibitem{zhang}
Lin-Bo Zhang, \emph{A parallel algorithm for adaptive local refinement of
  tetrahedral meshes using bisection}, Numer. Math. Theor. Meth. Appl.
  \textbf{2} (2009), no.~1, 65--89.

\end{thebibliography}

\end{document}